\documentclass[11pt]{article}                               

\usepackage{latexsym}
\usepackage{amssymb}

\def\AFOUR{%
\setlength{\textheight}{9.0in}%
\setlength{\textwidth}{5.75in}%
\setlength{\topmargin}{-0.375in}%
\hoffset=-.5in%
\renewcommand{\baselinestretch}{1.17}%
\setlength{\parskip}{6pt plus 2pt}%
}

\AFOUR                                           

\makeatletter
\def\section{\@startsection {section}{1}{\z@}{-3.5ex plus -1ex minus
 -.2ex}{2.3ex plus .2ex}{\large\sc}}
\def\subsection{\@startsection{subsection}{2}{\z@}{-3.25ex plus -1ex minus
 -.2ex}{1.5ex plus .2ex}{\normalsize\sc}}
\makeatother

\makeatletter
\@addtoreset{equation}{section}

\makeatother

\newcommand{\nc}{\newcommand}
\newcommand{\rnc}{\renewcommand}

\nc{\bea}{\begin{eqnarray}}
\nc{\eea}{\end{eqnarray}}
\nc{\be}{\bea}
\nc{\ee}{\eea}

\rnc{\a}{\alpha}
\nc{\ab}{\bar{\a}}
\nc{\ap}{\a^{+}}
\nc{\abm}{\ab^{-}}
\rnc{\b}{\beta}
\nc{\bb}{\bar{\b}}
\nc{\bbp}{\bb_{\zb}^{+}}
\nc{\bm}{\b_{z}^{-}}
\nc{\oa}{\overline{\a}}
\nc{\ob}{\overline{\b}}
\rnc{\gg}{\gamma}
\rnc{\d}{\delta}
\nc{\f}{\phi}
\nc{\fb}{\bar{\phi}}
\nc{\vf}{\varphi}
\nc{\p}{\psi}

\rnc{\c}{\chi}
\nc{\la}{\lambda}
\nc{\m}{\mu}
\nc{\n}{\nu}
\rnc{\o}{\omega}
\nc{\Om}{\Omega}
\rnc{\t}{\theta}
\nc{\eps}{\epsilon}
\rnc{\S}{\Sigma}
\nc{\F}{\Phi}

\nc{\trac}[2]{{\textstyle\frac{#1}{#2}}}

\nc{\ex}[1]{\mbox{e}^{\,\textstyle#1}}

\nc{\mat}[4]{\left(\begin{array}{cc}#1&#2\\#3&#4\end{array}\right)}

\nc{\som}[9]{\left(\begin{array}{ccc}#1&#2&#3\\#4&#5&#6\\#7&#8&#9%
\end{array}\right)}

\nc{\tr}{\mathop{\mbox{tr}}\nolimits}
\nc{\ad}{\mathop{\mbox{ad}}\nolimits}
\nc{\Tr}{\mathop{\mbox{Tr}}\nolimits}
\nc{\Det}{\mathop{\mbox{Det}}\nolimits}
\nc{\rk}{\mathop{\mbox{rk}}\nolimits}
\nc{\ra}{\rightarrow}
\nc{\Ra}{\Rightarrow}
\nc{\LRa}{\Leftrightarrow}
\nc{\ot}{\otimes}
\rnc{\ss}{\subset}
\nc{\nul}{\noindent\underline}
\nc{\non}{\nonumber\\}
\nc{\ZZ}{\mathbb{Z}}
\nc{\RR}{\mathbb{R}}
\nc{\CC}{\mathbb{C}}

\nc{\subs}[1]{{\vspace*{0.5cm}}%
{\noindent\underline{#1}}%
{\vspace*{0.3cm}}}

\nc{\zb}{\bar{z}}
\rnc{\lg}{\mathfrak{g}}
\nc{\lt}{\mathfrak{t}}
\nc{\lk}{\mathfrak{k}}
\nc{\lh}{\mathfrak{h}}
\nc{\pik}{\Pi_{\lk}}
\nc{\pip}{\Pi_{+}}
\nc{\pim}{\Pi_{-}}
\nc{\pih}{\Pi_{\lh}}
\nc{\jz}{J_{z}}
\nc{\jzh}{\jz^{\lh}}
\nc{\jzp}{\jz^{+}}
\nc{\jzm}{\jz^{-}}
\nc{\del}{\partial}
\nc{\dz}{\del_{z}}
\nc{\dzb}{\del_{\bar{z}}}
\nc{\az}{A_{z}}
\nc{\azb}{A_{\bar{z}}}
\nc{\g}{g^{-1}}
\nc{\dw}{\Delta_{W}}
\nc{\Ad}{{\mbox{Ad}}}
\nc{\ks}{Ka\-za\-ma-\-Su\-zu\-ki}
\nc{\KS}{\ks}
\nc{\ksm}{\ks\ model}
\rnc{\AA}{{\Bbb A}}
\nc{\BB}{{\Bbb B}}
\nc{\PP}{{\Bbb P}}
\nc{\cpm}{\CC\PP(m)}
\nc{\cpn}{\CC\PP(n)}
\nc{\cp}[1]{\CC\PP(#1)}
\nc{\gmn}{G(m,m+n)}
\nc{\gmnk}{\gmn_{k}}
\nc{\cO}{{\cal O}}
\nc{\bcO}{\bar{\cO}}
\nc{\bO}{\bar{O}}
\nc{\oQ}{\overline{Q}}

\begin{document}
\global\parskip=4pt

\begin{titlepage}
\begin{center}
\vspace*{.5in}
{\LARGE\bf Chern-Simons Theory on $S^1$-Bundles:}\\[.3in] 
{\Large\bf Abelianisation and q-deformed Yang-Mills Theory}\\
\vskip 0.7in
{\bf Matthias Blau}\footnote{e-mail: matthias.blau(at)unine.ch}
\vskip .1in
Institut de Physique, Universit\'{e} de Neuch\^{a}tel, Switzerland.
\vskip 0.2in
\vskip 0.2in
{\bf George Thompson}\footnote{e-mail: thompson(at)ictp.it}
\vskip .1in
ICTP, 
P.O. Box 586, 
34100 Trieste, 
Italy.\\

\end{center}
\vskip .4in
\begin{abstract} 
\noindent 
We study Chern-Simons theory on 3-manifolds $M$ that are circle-bundles
over 2-dimensional surfaces $\Sigma$ and show that the method of
Abelianisation, previously employed for trivial bundles $\Sigma \times
S^1$, can be adapted to this case. This reduces the non-Abelian theory on
$M$ to a 2-dimensional Abelian theory on $\Sigma$ which we identify with
q-deformed Yang-Mills theory, as anticipated by Vafa et al. We compare
and contrast our results with those obtained by Beasley and Witten using
the method of non-Abelian localisation, and determine the surgery and
framing presecription implicit in this path integral evaluation. We
also comment on the extension of these methods to BF theory and other
generalisations.
\end{abstract}

\vfill
\begin{center}
{\it Dedicated to Bob Delbourgo on the occassion of his 65th Birthday}
\end{center}
\vfill

\end{titlepage}


\begin{small}
\tableofcontents
\end{small}

\setcounter{footnote}{0}

\section{Introduction}

There has been a great success in computing the Chern-Simons partition
function by using conformal field theory techniques combined
with surgery. On the other hand there are very few exact (i.e.\
non-perturbative) gauge theory path integral computations of Chern-Simons
theory, the exceptions being on manifolds of the form $\Sigma_{g} \times
S^{1}$ \cite{btg/g}. The technique adopted in \cite{btg/g}, and reviewed
in \cite{btictp,btjmp}, was Abelianisation of the non-Abelian theory.

In this paper, we will study Chern-Simons theory on 3-manifolds $M_{(g,p)}$
that are non-trivial (monopole degree $p$) circle bundles over
2-dimensional (genus $g$) surfaces. Thus $M_{(g,0)}=\Sigma_g \times S^1$,
$M_{(0,p)}=L(p,1)$ are Lens spaces etc. 

There are three principal reasons for us for looking at this issue:

\begin{itemize}

\item From a path integral technological point of view, the question 
arises, if the method of Abelianisation can be generalised from trivial
($p=0$) 
to non-trivial circle bundles: Abelianisation works well in two dimensions;
thus one needs to be able to ``push down'' things from the 3-manifold to the
base, which is a somewhat less obvious procedure 
in the case of non-trivial bundles.

\item Chern-Simons theory on the 3-manifolds $M_{(g,p)}$
(and more general Seifert
manifolds) has recently been studied by Beasley and Witten \cite{BW}
using the method
of non-Abelian localisation \cite{ewym}. In this context the question arises
whether the diagonalisation procedure (once one has established that it is
applicable) yields results that are manifestly equivalent or 
comparable to those of \cite{BW} and if there are situations in which one or
the other method is more efficient.

\item Chern-Simons theory on Lens spaces $L(p,1)=M_{(0,p)}$ 
has also recently appeared in the context of black hole partition
function calculations via topological string theory \cite{vafaym,AOSV}. There,
methods of Abeliansiation were used to argue that this theory is equivalent 
to a ``q-deformed'' two-dimensional Yang-Mills theory. In \cite{AOSV} the
connection with Chern-Simons theory was somewhat indirect and the question
arises if it is possible to derive the relation between the action of
Chern-Simons theory and that of a two-dimensional action
in a direct manner.

\end{itemize} 

The Chern-Simons action is
\be
kS_{CS}[A]=\frac{k}{4\pi}\int_{M} \Tr \left( AdA + \frac{2}{3} A^{3} \right)
\;\;.
\ee
In order to get a handle on the cubic part of the Chern-Simons action,
we make use of the geometry of $M_{(g,p)}$, following \cite{BW}, to
decompose the connection into a horizontal and a vertical part, which
now appear at most quadratically in the action and thus lend themselves
to a path integral treatment.\footnote{In contrast to \cite{BW}, however,
we do not introduce a corresponding St\"uckelberg field and shift symmetry
associated with this decomposition.}

In particular, due to the non-triviality of the bunlde one finds a
term quadratic in the vertical component of the connection (a scalar
$\phi$ from the point of view of the base) from the quadratic term of
the Chern-Simons action, suggesting, already at this stage, a relation
with some kind of 2-dimensional Yang-Mills theory rather than with a
BF- or $G/G$-like theory (as encountered for $p=0$ in \cite{btg/g}).
This new term apart, the action resembles that of Chern-Simons theory
on a trivial bundle $\Sigma_g \times S^1$, and we can now attempt to apply
the methods of \cite{btg/g} to this case.

We can summarise the results that we find as follows:

\subs{Diagonalisation}

The method of diagonalisation is applicable to the case of non-trivial
circle bundles and permits one to reduce the non-Abelian 3-dimensional
Chern-Simons theory to a 2-dimensional Abelian theory (whose partition
function can in many cases be evaluated in a straightforward manner).

The main differences to the previously discussed case of trivial $S^1$-
bundles $M_{(g,0)}$ are:

\begin{itemize}

\item the obstruction bundles to diagonalisation (which the method of
Abelianisation instructs one to sum over in the path integral) are now
precisely the torsion bundles on the 3-manifold $M_{(g,p)}$.

\item the fields that one integrates out, in the process of reducing the 
theory from 3 to 2 dimensions (the non-$U(1)$-invariant fields), 
are now sections of non-trivial bundles $\mathcal{O}(-np)$
on the base $\Sigma_g$; 
correspondingly this changes the evaluation of the determinants.

\end{itemize}

The abelianised expression for the partition function $Z_k$ of the level $k$
Chern-Simons theory with gauge group $G$ we find is\footnote{For $G$ 
simply-laced and simply-connected; in this introductory section we will also
suppress certain overall normalisation and phase factors from the equations.}  
\be
Z_k[M_{(g,p)},G]
= \sum_{r \in \mathbb{Z}_{p}^{\mathbf{rk}}} \,  \int_{\lt}
T_{S^{1}}(\phi)^{\chi(\Sigma_{g})/2}\, \exp{i \frac{k+
      c_{\lg}}{4\pi} \Tr \left( p \, \phi^{2} + 4 \pi r\, \phi
  \right)}
\label{zkmg}
\ee
The integral over the Cartan subalgebra $\lt$ 
of $G$ is all that remains
of the integral over the vertical component of the gauge field, the integral
over all other modes and fields, in particular the horizontal components of
the gauge fields, having already been performed. 
Here $\mathbf{rk}=\dim\lt$ is the rank of $G$ and $\chi(\Sigma_g)=2-2g$ the
Euler characteristic of $\Sigma_g$.
The Ray-Singer torsion $T_{S^1}$ of $S^1$ and the shift $k \ra k+
c_{\lg}$ arise from the absolute value and phase of the ratio of determinants
generated by the integral over these gauge field modes and the ghosts. 
The sum over $r$ is the sum over the torsion classes of line bundles 
mentioned above. 

\subs{Comparison with non-Abelian Localisation} 

In a recent paper, Beasley and Witten \cite{BW} adapted the method of
non-Abelian localisation (originally developed for 2-dimensional Yang-Mills
theory in \cite{ewym}) to Chern-Simons theory on Seifert manifolds.

The upshot of this localisation is that the partition function can
schematically be written as sums of integrals of the form
\be
Z_k[M,G]=\sum\int_{\lg \times \mathcal{M}} \mathcal{F} = \sum\int_{\lt \times
  \mathcal{M}} \widehat{\mathcal{F}} \label{bw}
\ee
for some integrand $\mathcal{F}$. Here $\lg$ is the Lie algebra of $G$ and
$\mathcal{M}$ is the component of the  moduli space (of flat or Yang-Mills
connections, say) onto which 
the theory localised, and the right hand side is what one gets on
applying the Weyl integral formula to reduce the Lie algebra integral
to the Cartan subalgebra $\lt$. 

One of the motivations for \cite{BW} was to explain formulae of
Rozansky and Lawrence \cite{RL} in the $SU(2)$ case. Following
previous work of Rozansky, Lawrence and Rozansky begin with the
conformal field theory formula for the Chern-Simons partition
function in terms of characters of integrable representations of
$SU(2)$. The empirical
discovery here was that the Chern-Simons partition function could be
expressed as coming from stationary phase contributions to the
path integral, that is as integrals and residues over the moduli space
of flat connections on $M$. The localisation of \cite{BW} explains in
an a priori manner why this is so.

A prototypical example are the Lens spaces $L(p,1)$. For
$G=SU(2)$, the formula of \cite{RL} is, in the normalisation of \cite{BW},
\be
Z_{k}[M_{(0,p)},SU(2)] \sim \sum_{r=0}^{p-1} \frac{1}{2\pi i}
\int_{C^{(r)}} dz\, (2\sinh{\frac{z}{2}})^{2}\, \exp{\left(
    \frac{i(k+2)}{8\pi}pz^{2} 
  - (k+2)rz\right) }\label{rl}
\ee
(we have suppressed an overall framing dependent phase)
where the contour $C^{(0)}= e^{\frac{i \pi}{4}} \times \mathbb{R}$ and
the other contours $C^{(r)}$ are parallel to this one through the
stationary phase point $z = -4\pi i r/p$. This is of the general form
(\ref{bw}), the sum being a sum over the flat connections on $L(p,1)$.
For the generalisation of such formulae to $SU(n)$, $n>2$, see \cite{marino}.

The formulae that we derive, on the other hand, take the general form
(cf.\ (\ref{zkmg}))
\be
Z_k[M_{(g,p)},G]=\sum\int_{\lt} \widetilde{\mathcal{F}}.
\label{bt}
\ee
This agrees with (\ref{bw}) when $\mathcal{M}$ is a point or a finite
union of points. In particular, (\ref{rl}) is also of the general
form (\ref{bt}) and it is easy to see (using the explicit form of the
Ray-Singer torsion and analytic continuation - there are no poles to
worry about) that our general result (\ref{zkmg}) reproduces (\ref{rl}).

More generally, however, a striking difference between (\ref{bw})
and (\ref{bt}) is the absence of an integration over $\mathcal{M}$
from the latter. This is a general feature of our strategy for solving
low-dimensional gauge theories, e.g. BF theory, via path integrals. In
the latter case, by integrating over $B$ first, one sees that one is
left with an integral over the moduli space of flat connections and is
essentialy calculating its volume. This integral is typically difficult
to perform in practice. By reversing the order of integration, however,
i.e.\ integrating first over all connections and only then over $B$, one
completely side-steps the issue of having to integrate over the moduli
space of flat connections and obtains explicit expressions for the
volume - see e.g.\ \cite{btictp}. Adopting the same strategy here,
we arrive at (\ref{zkmg}), which is of the general form (\ref{bt}),
with the integral over $\mathcal{M}$ having, somewhat miraculously,
been taken care of. Since the moduli spaces in question can be quite
nasty and singular, not having to work with them directly is a blessing.

\subs{Relation with qYM-Theory}

An intermediate step in arriving at the final formula (\ref{zkmg}) is 
the reduction of Chern-Simons theory on $M_{(g,p)}$ to an effective
2-dimensional Abelian theory. This theory, given by (\ref{2dpi}), with
action (\ref{ssigma}),
\be
S_{\Sigma}[A_H,\phi]= \frac{k+ c_{\lg}}{4\pi} \int_{\Sigma_g} \Tr \, 
(2\phi \, F_{H} + p \phi^{2} \omega)\;\;,
\ee
and the gauge field integration range as specified in (\ref{2dfh}), is of
the abelianised BF- or $G/G$-model type with an additional $\phi^2$-interaction.
This theory can be considered as a deformation of ordinary
Yang-Mills theory, the deformation residing in the measure and the finite
sum over torsion bundles. Since the theory that one obtains for $p=0$,
namely the $G/G$-model and its Abelianisation, can be interpreted as
a q-deformed BF theory (in the sense that in the expression for the
partition function dimensions of representations are replaced by their
quantum-dimensions - see e.g.\ \cite{btg/g}), it is natural to suspect
that what one will find for $p\neq 0$ is a corresponding q-deformation
of Yang-Mills theory.

In \cite{AOSV}, on the other hand, an alternative deformed 
2-dimensional Yang-Mills description
of Chern-Simons theory on Lens spaces $M_{(0,p)}$ was proposed, involving an
action of the above form, but with a compact scalar 
(so that the $\phi^2$-term in the action requires some interpretation) 
and an infinite sum over torus bundles, both in apparent contrast with 
(\ref{zkmg}, \ref{rl}).  
The partition function of this theory for general $M_{(g,p)}$ is 
\be
\tilde{Z}_k[M_{(g,p)},G]= \sum_{r \in \mathbb{Z}^{\mathbf{rk}}} \,
\int_{\lt/I} 
T_{S^{1}}(\phi)^{\chi(\Sigma_{g})/2}\, \exp{i \frac{k+
      c_{\lg}}{4\pi} \Tr \left( p \, \phi^{2} + 4\pi r\, \phi
  \right)}
\label{zqym}
\ee
where $I$ is the integral lattice (so that $\phi$ is now compact) and the 
sum over line bundles is not constrained. It was shown in \cite{AOSV} that 
this partition function bears the same relation to q-deformed representation
theory as ordinary Yang-Mills theory does to ordinary representation theory,
and thus the theory can legitimately be referred to as q-deformed Yang-Mills
theory.

We will show in section 6 that, despite
appearance, (\ref{zqym}) agrees precisely with our result (\ref{zkmg}), 
\be
\tilde{Z}_k[M_{(g,p)},G]= Z_k[M_{(g,p)},G]\;\;,
\ee
so that the q-deformation can be equivalently regarded as arising from either a
compact scalar or a restricted sum over torus bundles, the latter description
arising more naturally from the point of view of diagonalisation.

In particular, this shows that the partition function on $M_{(g,p)}$ is the
same as the expectation value of an operator in the $G/G$-model, i.e.\ in
the theory on $M_{(g,0)}=\Sigma_g \times S^1$ where $\phi$ is naturally
compact due to large gauge transformations. Concretely, up to a phase 
proportional to $p$ (which we calculate), one has
\be
Z_k[M_{(g,p)}, G ] = \langle \ex{ip\frac{(k+c_{\lg})}{4\pi} \Tr 
  \phi^{2} }
\rangle_{\Sigma_{g} \times S^{1}} \;\;.
\label{insertion}
\ee
This means that manifolds of non-trivial Chern classes are simply
created by insertions of the operator $\exp{i\frac{(k+c_{\lg})}{4\pi}
\Tr \phi^{2} }$ in the path integral for the trivial bundle.  This has
been argued before by Vafa \cite{vafaym} in a rather different way.
Our calculation provides an \textit{ab initio} path integral derivation 
of this cute fact.

For earlier work on q-deformed Yang-Mills theory see \cite{roche}
and, in particular, \cite{klimcik} where a Lagrangian realisation of
this theory was proposed and solved by the method of diagonalisation.
Aspects of the relation between Chern-Simons theory on Lens spaces and
Yang-Mills theory were previously discussed in \cite{deharo}. For other
recent work on q-deformed Yang-Mills theory see \cite{qymother,deharo2}.

\subs{Comparison with the Surgery Prescription}

The explicit formulae that we obtain for the partition function in section
\ref{partfm} have the form that one would find on performing surgery on
knots and links in $\Sigma_{g} \times S^{1}$.  Indeed, keeping track
of all phases, the partition function can be written in terms of the
standard modular $S$- and $T$-matrices of the Wess-Zumino-Witten model as
\be
Z_k[M_{(g,p)}, G] = \sum_{\lambda} S_{0\lambda}^{2-2g} \,
T_{\lambda}^{-p}\;\;,
\ee
where the sum is over level $k$ integrable weights.

Quite generally, starting from $M_{(g,0)}$ different surgeries can yield
the same 3-manifold $M_{(g,p)}$ but this manifold will come equipped with
a framing which depends on the surgery so the formulae we obtain involve
an implicit choice of framings. The evaluation of the path integral then
is always in some framing of the 3-manifold (and links) in question.
However, it is not at all transparent from the outset which framing one
is actually in.

{}From their results, Beasley and Witten \cite{BW} deduce that,
in their calculations, they are in the `Seifert' framing and not in the
canonical framing of the 3-manifold. Starting from $S^{2}\times S^{1}$ one
generates $S^{3}$, by acting with $T^{m}ST^{n}$.  The canonical framing
for $S^{3}$ corresponds to $m=n=0$, while one obtains a $U(1)$-invariant
`Seifert' framing for $n+m =2$.

Our previous calculations \cite{btg/g} were in the canonical framing for
$M_{(g,0)}$, so one can ask which surgery prescription is being used to
generate the $M_{(g,p)}$ with the framing that is employed in our path
integral. By considering only the partition function we guess that the
surgery prescription we are implicitly using to get from $M_{(g,0)}$
to $M_{(g,p)}$ is to act with $(TST)^{p}$.

The surgery prescription is always such that the expectation value of
the Hopf link equals the surgery matrix. In \cite{btg/g} Abelianisation
was employed to also compute expectation values of Wilson lines in the
$S^{1}$ direction of $M_{(g,0)}$. In section \ref{wl} we extend this to
compute the expectation values of Wilson lines in the non-trivial fibre
direction of $M_{(g,p)}$. In particular, we show that the expectation
value of the Hopf link is indeed $TST$, confirming the guess that we made.

\subs{Generalisations}

We have not aimed for maximal generality in this article, and the results
can be generalised in various ways, to other three-manifolds, other groups,
and other 3-dimensional gauge theories (such as BF theory). We will briefly
come back to these issues in section 9.

\section{Gauge Theories on 3-Manifolds $M_{(g,p)}$}

We will consider gauge theories (and later on more specifically
Chern-Simons theory) on 3-manifolds $M_{(g,p)}$ which are
themselves principal $U(1)$ bundles $U(1) \rightarrow M_{(g,p)}
\stackrel{\pi}{\rightarrow} \Sigma_{g}$ over 2-dimensional surfaces
$\Sigma_g$ of genus $g$ and first Chern (or Euler) class $-p\in\mathbb{Z}$
(under the identification $H^2(\Sigma_g,\ZZ)=\ZZ$; the minus sign is
a consequence of our conventions which we spell out below).

We choose the gauge group $G$ to be compact, (semi-)simple, and simply
connected. In particular, this implies that a principal $G$-bundle on a
3-manifold $M$ and all its associated vector bundles are trivial.  In this
case we understand the global obstructions to diagonalisation well enough
to be able to apply this method to gauge theories on $M$. For comments
on the more general case we refer to \cite{btdiag} and the discussion in
section 9.  We also assume, for convenience and notational simplicity, that
$G$ is simply laced even though this latter assumption is not necessary.

The analysis of gauge theories on such 3-manifolds $M_{(g,p)}$
is greatly simplified
by appropriate gauge choices that are adapted to the geometry of the
situation at hand. In the case $p=0$, i.e.\ $M=M_{(g,0)}=\Sigma_g
\times S^1$ \cite{btg/g}, it was natural to single out the ``vertical''
component of the connection, $A_{\theta}$ say, with $\theta$ an angular
fibre coordinate, and to impose, as a first step, the condition
\be
\del_{\theta}A_{\theta}=0
\label{gc1}
\ee
(the simpler axial gauge condition $A_{\theta}=0$ not
being available because of the possibility of having non-trivial holonomy
along the $S^1$). In a second step, it was then possible (and very effective)
to use the residual 2-dimensional gauge invariance to ``diagonalise''
$A_{\theta}$, i.e.\ to conjugate it into the Cartan subalgebra $\lt$ of the
Lie algebra $\lg$ of $G$. This is tantamount to imposing the condition
\be
A_{\theta}^{\lk}=0
\label{gc2}
\ee
where $\lk$ is the orthogonal complement of $\lt$ in $\lg$, 
$\lg = \lt \oplus \lk$, with respect to the Killing-Cartan form.

In order to mimic this gauge fixing procedure for $p\neq 0$,
we want to again single out out one particular component
of the gauge field. This amounts to picking a one dimensional sub-bundle of
the cotangent bundle $T^{*}M$ of $M$. For $M=M_{(g,p)}$ there is a natural
way to do this.

\subs{Geometric Set-Up}

Indeed, let $\kappa$ be a connection on the principal $U(1)$-bundle
$M_{(g,p)}$, thought of as a globally defined real-valued 1-form on the 
total space of the bundle, 
and denote by $K$ the fundamental vector field on $M_{(g,p)}$, i.e.\ the
generator of the $U(1)$-action. A connection $\kappa$ is characterised
by 
\be
\iota_{K}\kappa = 1 \label{inner}
\ee
and the equivariance condition
\be
L_{K} \, \kappa = 0\;\;,
\label{equiv}
\ee
where $L_{K} = \{ d\, , \, \iota_{K} \}$ is the Lie derivative in the
$K$ direction. These two conditions imply that $\iota_K d\kappa=0$, i.e.\
the expected statement 
that the curvature 2-form $d\kappa$ of $\kappa$ is horizontal.

Note that $\kappa$ is not unique. In local coordinates one has
\be
\kappa = d\theta + a\;\;,\label{kap}
\ee
where $\theta$ is a fibre coordinate, $0\leq \theta <1$, and
$a=a_{i}\, dx^i$ is a local representative on $\Sigma_g$ of the
connection $\kappa$ on $M_{(g,p)}$. Since $M_{(g,p)}$ has degree $p$,
we may choose $\kappa$ (and hence $a$) so that the curvature 2-form satisfies
\be
d\kappa = p\, \pi^{*}(\omega)
\label{pomega}
\ee
for $\omega$ a unit normalised symplectic form on $\Sigma_{g}$, 
\be
\int_{\Sigma_{g}} \omega =1.
\label{omega}
\ee

In passing we note that for $p\neq 0$
a choice of $\kappa$ equips $M_{(g,p)}$ with what is
known as a contact structure, i.e.\ a 
1-dimensional sub-bundle of $T^{*}M$ (generated by a 1-form $\kappa$)
such that $\kappa \wedge d\kappa$ is nowhere vanishing on $M$. 
Indeed, with the above choices we see that
\be
\kappa \wedge d \kappa = p\, d\theta \wedge 
\pi^{*}(\omega) 
\ee
is nowhere vanishing as required providing that the $U(1)$ bundle is
non-trivial, that is providing $p\neq 0$. For later use we note that
\be
\int_{M} \kappa \wedge d \kappa =  p \int_{\Sigma_{g}} \omega = p\;\;.\label{p}
\ee
Thus, depending on the sign of $p$, $\int_M \kappa\wedge d\kappa$ may be
either positive or negative. In this respect our conventions differ from
those of \cite{BW} where the orientiation of $M$ is chosen such that $p$
is non-negative.
 
Contact structures can be put on any compact orientable 3-manifold
\cite{M} but, as we have seen, are particularly simple to describe
when $M$ is a principal $U(1)$-bundle.  This contact structure point
of view played an important role in the considerations of \cite{BW}.
In the present paper, however, we will downplay the role of the contact
structure somewhat in order to bring out the analogy with the case $p=0$
for which the above construction fails to provide a contact structure.

Our convention for $U(1)$ is that the generator of its Lie algebra is
$i$. A connection on a $U(1)$ bundle is locally 
\be
\varpi =2\pi i\kappa = 2\pi i (a +  d\theta)\;\;.
\ee
Chern classes are generated by $\det{(I - \frac{1}{2\pi i }d\varpi)} $ so
that $c_{1} = [-d\kappa]$. Consequently the first Chern class of
the naturally associated line bundle to $M_{(g,p)}$ is
\be
-\int_{\Sigma_{g}} d \kappa = -p\;\;.\label{fcc}
\ee
Comparing with the discussion in \cite[p.121]{BottTu}, we see that
$\kappa$ is precisely what is called an angular form there (and denoted
by $\psi$): this form has the property that the vertical component is the
unit volume form, with the standard orientation, and that its exterior
derivative is minus the pull back of the Euler class, or first Chern
class, of the associated bundle; $\kappa$ has these properties since
its vertical component is $d\theta$ and $d\kappa = - \pi^{*}(-p \omega)$.

\subs{Decomposition}

As both $K$ and $\kappa$ are 
nowhere vanishing we have, by virtue of (\ref{inner}), that $\kappa
\wedge \iota_K$ and $(1-\kappa \wedge \iota_K)$ are projection operators, 
corresponding to  the decomposition
\be
T^{*}M = T^{*}_{\kappa}(M)\oplus T^{*}_{H}(M), \;\;\;
T^{*}_{\kappa}(M) \approx  \Omega^{0}(M) 
\ee
into forms along the $\kappa$ direction and those which are
horizontal. Concretely, for 
$\alpha \in
\Omega^{1}(M, \mathbb{R})$ one has $\alpha = \alpha_{\kappa} + \alpha_{H}$ with 
\be
\alpha_{\kappa}=\kappa \wedge \iota_K \, \alpha \in \Omega^{1}_{\kappa}(M,
\mathbb{R}),
\;\;\; \alpha_{H} = (1-\kappa
\wedge \iota_K) \, \alpha \in \Omega^{1}_{H}(M,\mathbb{R}).
\ee
Likewise we can decompose connections on vector bundles $E$ over $M$, 
thought of as elements of $\Omega^{1}(M, \lg)$, 
\be
A = A_{\kappa} + A_{H}\equiv  \phi \, \kappa +
A_{H}. \label{decomp} 
\ee
Since $\phi \in \Omega^{0}(M, \lg)$ we can think of it as a section of the
adjoint bundle $E=M \times \lg$. Its transformation behaviour under
infinitesimal gauge transformations $\d A = d_A \Lambda$ is 
\be
\d \phi = \mathcal{L}_{\phi}\Lambda\;\;,
\ee
with
\be
\mathcal{L}_{\phi}  = L_{K} + [\phi, \; ].
\ee

There is also a decomposition of the exterior derivative
\be
d = (1-\kappa \wedge \iota_K)d + \kappa \wedge \iota_K 
d \equiv \pi^* d_{\Sigma} 
+ d_{K}.
\ee
On horizontal forms $B_{H}$, $\iota_K B_{H}=0$, one has $\iota_K\, dB_{H} =
L_{K}\, B_{H}$, with  
$L_{K}\, B_{H}$ also horizontal, and therefore
\be
dB_{H} = (\pi^*d_{\Sigma}) B_{H} + \kappa \wedge L_{K}\, B_{H} \;\;,
\ee
and, in particular,
\be
B_{H}\wedge dB_{H} = B_{H} \wedge \kappa \wedge L_{K}\, B_{H}\;\;.
\ee
{}From (\ref{equiv}) we also have the useful fact
\be
\kappa \wedge L_{K}\, B_{H} = L_{K} (\kappa \wedge B_{H}).
\ee

\subs{Gauge Choices}

Having singled out a particular component of the gauge field
$A_{\kappa}$ it is tempting to impose the gauge
condition $A_{\kappa} = 0 = \phi$. However, just as for $p=0$,
this is not possible since Wilson loops along the fibres of $M_{(g,p)}\ra
\Sigma_g $ are gauge invariant and non-trivial
(we will discuss their correlation functions in Chern-Simons theory in 
section 8).

Instead we may (and do) impose the analogue of the condition (\ref{gc1}),
namely 
\be
L_K A_{\kappa}=0 \Leftrightarrow 
L_{K}\phi = \iota_{K} \, d \,\phi = 0\;\;.
\label{gf}
\ee
This gauge condition, $L_{K}\phi=0$, tells us
that $\phi$ is a $U(1)$-invariant section of $E$. Equivalently, it can 
therefore be regarded as a section of the (trivial) adjoint bundle $V$
over $\Sigma_g$ (see Appendix \ref{apgysin}).

Having pushed down $\phi$ to $\Sigma_g$ in this manner, we can now
prooceed to the diagonalisation of $\phi$ as in \cite{btg/g}. Thus
let $T$ be some maximal torus of $G$ and $\lt$ the corresponding Cartan
subalgebra, with  $\lg = \lt \oplus \lk$. We now impose the analogue of
(\ref{gc2}), namely
\be
\phi^{\lk} = 0. \label{gfc}
\ee
As shown in \cite{btg/g,btdiag}, the price for diagonalising sections
of $V$ is that in the path integral, 
when we come to it, we must sum over all $T$-bundles on $\Sigma_g$,
hence from the 3-dimensional perspective over all $T$-bundles 
that one gets by pull back from $\Sigma_{g}$. 
Since the pull-back $\pi^*M_{(g,p)}$ of the $U(1)$-bundle $M_{(g,p)} 
\stackrel{\pi}{\rightarrow} \Sigma_{g}$ to the total space $M_{(g,p)}$ is
(tautologically) trivial, $\pi^*M_{(g,p)}=M_{(g,p)}\times U(1)$, the pull-back of
the $p$-th power of any line bundle on $\Sigma_g$ to $M_{(g,p)}$ is trivial.
Thus the pull-backs of line bundles from $\Sigma_g$ to $M_{(g,p)}$ are of
finite order.  We show in Appendix \ref{apgysin} that all torsion
(finite order $p$) 
bundles on $M_{(g,p)}$ arise in this way, so that it is precisely these
bundles that we should sum over in the path integral.

\subs{Ghost Action}

The BRST symmetry of the gauge theory is standard and we do not repeat
it here. We mimic arguments presented in detail in \cite{btg/g} for
fixing the gauge and the associated ghost terms.
Both of the conditions (\ref{gf}) and (\ref{gfc}) can be
simultaneously imposed by adding the BRST exact terms
\be
\int_{M} \left[ E* \phi + \overline{c}*\,  \mathcal{L}_{\phi} \, c\,
\right] = \int_{M} \left[ E* \phi + d \kappa \wedge \kappa\,\overline{c}\,
  \mathcal{L}_{\phi} \, c\,
\right]
\label{ghost} 
\ee
with the understanding that those modes which are $U(1)$-invariant,
i.e. solutions to the equations 
\be
L_{K}\, E^{\lt} = L_{K}\, c^{\lt} = L_{K}\, \overline{c}^{\lt} = 0\;\;,
\ee
are not to be included in the path integral.

Here $*$ refers to a metric $g_M$ on $M$. It is convenient to choose this
metric to be $U(1)$-invariant, and a natural choice (which we will adopt) is
\be
g_M = \pi^*g_\Sigma + \kappa \otimes \kappa
\label{gmgs}
\ee
with $g_\Sigma$ a metric on $\Sigma_g$ such that $*_{\Sigma}1 = \omega$
(\ref{omega}).  

Thus far the discussion has not been theory specific. All of the
considerations above could be applied to, say, Yang-Mills theory on
$M$. In the next section we move on to the theory of interest for us.

\section{Chern-Simons Theory on  $M_{(g,p)}$}\label{CSonM}

The level $k$ Chern-Simons action is
\be
kS_{CS}[A]=\frac{k}{4\pi}\int_{M} \Tr \left( AdA + \frac{2}{3} A^{3} \right)
\ee
In terms of the decomposition (\ref{decomp}) the integrand becomes
\be
\Tr \left( A_{H}\wedge d_{\phi}\, A_{H} +  \phi\,\kappa\wedge  d\, A_{H} +
\phi\, d\kappa\wedge A_{H} 
 + \kappa \wedge A_{H}\wedge d \phi +
 \phi^{2}\, \kappa\wedge d \kappa \right) \label{csf1}
\ee
where
\be
d_{\phi} = d  + \kappa \wedge [ \phi, \; ]\;\; .
\ee
Note in particular the appearance of a term quadratic in $\phi$ for $p\neq
0$.

The $G$-bundles we are considering are trivial so we may take $A$ to
be a Lie algebra valued form. This means that, up to a total derivative,
we can rewrite (\ref{csf1}) as
\be
\Tr \left(A_{H}\wedge d_{\phi} \, A_{H} + 2\phi \, \kappa \wedge d\, A_{H} +
\phi^{2}\, \kappa\wedge d \, \kappa \right) \;\;.\label{csf}
\ee

Since the forms $A_{H}$ are orthogonal to $\kappa$, the first term
necessarily only involves a derivative in the direction of 
$\kappa$,
\be
\Tr \, A_{H}\wedge d_{\phi} A_{H} = \Tr \, A_{H} \wedge \kappa\wedge
\mathcal{L}_{\phi} \, A_{H} 
\label{kin1}
\ee
(while the derivative in the second term
acts only in the horizontal direction)
and thus we can write the action as
\be
kS_{CS}[A_H,\phi] = \frac{k}{4\pi}\int_M \Tr\left(
A_{H}\wedge \kappa \wedge \mathcal{L}_{\phi}
\, A_{H} + 2\phi \, \kappa \wedge d\, A_{H} +
\phi^{2}\, \kappa\wedge d \, \kappa \right)\;\;.
\label{sm0}
\ee

\subs{Conditions on $\phi$}

Consider those $A_{H}^{\lt}$ which are $U(1)$ invariant,
\be
L_{K} \, A_{H}^{\lt}=0\;\;.
\ee
It follows from (\ref{kin1}) that 
these fields do not appear in the kinetic term $A_{H}\wedge  d_{\phi}
A_{H}$.  Consequently they
only appear in the mixed kinetic term $2\phi \, \kappa \wedge  d\,
A_{H}$. The path 
integral over such $A_{H}^{\lt}$ then imposes a (delta function) condition
on $\phi$, namely
\be
\iota_{K}d (\kappa \, \phi) = 0\;\;.
\ee
This delta function constraint on $\phi$ together with the gauge
condition (\ref{gf}) imply that $\phi$ is actually constant,
\be
d \phi =0.
\ee
We will come back to this argument and its consequences in section 5.
Finally, with $\phi$ constant we have, from (\ref{p}), that
\be
\int_{M} \Tr \, \kappa \wedge d\kappa \, \phi^{2} =  p \,\Tr \, \phi^{2}
\ee

\section{Reduction to an Abelian Theory on $\Sigma_{g}$}\label{reduction}

Having discussed the effect of integrating out the $U(1)$-invariant modes
of $A_{H}^\lt$, we now keep these and investigate what happens upon 
integrating out the other modes and fields, with the understanding that
$\phi$ will ultimately turn out to be constant. All these fields appear
quadratically in the action, and therefore will give rise to ratios of
determinants (whose definition and regularisation we will subsequently 
discuss in detail in Appendix \ref{regs}).

Given the choice of metric (\ref{gmgs}), the operator $*\,\kappa \wedge \,
\mathcal{L}_{\phi}$ acts on the space 
of  horizontal $\lk$-valued 1-forms,
\be
*\, \kappa \wedge \,  \mathcal{L}_{\phi}:\Omega^{1}_{H}(M, \lk)
\rightarrow \Omega^{1}_{H}(M, \lk) .
\ee
Hence integrating over the $\lk$-components of the ghosts ghosts $(c^{\lk},
\overline{c}^{\lk})$ and the connection $A_{H}^{\lk}$, one obtains the
following ratio of determinants:
\be
\frac{\Det{ \left(
      i\mathcal{L}_{\phi}\right)_{\Omega^{0}(M, \lk)}
  }}{\sqrt{\Det{\left(*\,  \kappa \wedge \,
       i \mathcal{L}_{\phi}\right)_{\Omega^{1}_{H}(M, 
      \lk)} } }}\;\;. \label{ratio}
\ee
Note that
\be
* \, \kappa = -  \iota_{K} \, * =  \star_{2}
\ee
where $\star_2$, introduced in \cite{BW}, is a lift of the Hodge duality
operator $*_\Sigma$ to $M_{(g,p)}$ in the sense that $\star_2  \pi^*=
\pi^* *_\Sigma$.  This Hodge operator therefore appears naturally in our
evaluation of the path integral and the definition of the determinants
(Appendix \ref{regs}).

Integration over the ghosts $(c^{\lt}, \overline{c}^{\lt})$ and those
$A^{\lt}_{H}$ modes which are 
not $U(1)$ invariant give the following ratio of determinants:
\be
\frac{\Det^\prime{ \left(
      iL_{K}\right)_{\Omega^{0}(M, \lt)}
  }}{\sqrt{\Det^\prime{\left(*\,  \kappa \wedge \,
        iL_{K}\right)_{\Omega^{1}_{H}(M, 
      \lt)} } }} \label{ratio2}
\ee
The notation $\Det^\prime$ indicates that the zero
mode of the operator is not included.

On integrating out all the $\lk$-valued fields as well as all the
$\lt$-valued modes which are not $U(1)$ invariant, the
Chern-Simons path integral essentially reduces to the 
path integral of an Abelian 2-dimensional gauge theory on
$\Sigma_{g}$. Assembling all the ingredients, this path integral is
\be
Z_k[M_{(g,p)},G]\sim 
\ex{4\pi i p\Phi_{0}}
\int D\phi\, DA^{\lt}_{H}\; T_{S^{1}}(\phi)^{\chi(\Sigma_{g})/2} \;
\exp{\left(i\, \frac{k+ c_{\lg}}{4\pi} \, S_{M}\right) }\label{2dpi}
\;\;.
\ee
The action is
\be
S_{M} = \int_{M} \Tr \, (2\phi \,
    \kappa  \wedge F_{H} + 
    \phi^{2} \, \kappa \wedge d\kappa ), 
\label{sm}
\ee
and since the path integral is only over invariant modes,
we can push the action $S_M$ down to
$\Sigma_{g}$. Explicitly, the 2-dimensional Abelian action reads, 
recalling (\ref{pomega}), 
\be
S_M\ra S_{\Sigma}[A_H,\phi]= \frac{k+ c_{\lg}}{4\pi} \int_{\Sigma_g} \Tr \, 
(2\phi \, F_{H} + p \phi^{2} \omega)\;\;,
\label{ssigma}
\ee
where $A_H=A_H^{\lt}$ and $\phi=\phi^{\lt}$. 

The various new terms appearing in (\ref{2dpi}) arise as follows: 

The ratios of determinants (\ref{ratio}, \ref{ratio2}) that appear
are almost unity. In calculating these ratios of determinants we
pay attention to the absolute value and to the phase.  As far as the
absolute value is concerned, the deviation from unity is due to the
mismatch in zero modes. This mismatch is just the Euler characteristic
$\chi(\Sigma_{g})$. The Euler characteristic appears from an index theorem
when we regularise as in \cite{btg/g} with the $\zeta$-function associated
to the Dolbeault operator, which is rather natural given the appearance
of the complex structure $\star_{2}$.

In this way, for the absolute value of the determinants one finds
$T_{S^{1}}(\phi)^{\chi(\Sigma_{g})/2}$ where 
\be
T_{S^{1}}(\phi) = \det{}_{\lk}{(1- \Ad{ \, \ex{\phi}})}
\label{ts1}
\ee
is the Ray-Singer torsion of $S^{1}$ (with
respect to the flat connection $2\pi i\phi d\theta$). 

On the other hand, when it comes to the phase, one does an $\eta$-function
calculation. This calculation gives us the the famous shift in the
level, $k \rightarrow k+ c_{\lg}$ as well as the (framing
dependent) phase $4\pi p\Phi_0$ where 
\be
\Phi_0 = \frac{1}{48}\dim G\;\;.
\ee

There are some other things about (\ref{2dpi}) that require comment.

The first is that, in writing (\ref{2dpi}) and (\ref{ts1})
we have not kept track of
the overall real normalisation of the path integral (while we have
kept track of the phase). For instance, in the path integral we should
also integrate over harmonic $A^{\lt}_{H}$-modes even though these do
not appear in the action. Fortunately, such modes are compact thanks
to the residual Abelian gauge symmetry, as explained in \cite{btg/g},
and consequently those modes give a finite volume factor to the path
integral.  We will fix the remaining real normalisation constant in
section \ref{partfm} by comparison with the known normalisation for $p=0$
\cite{btg/g}.

The second comment is that (\ref{2dpi}) as it stands is incomplete as
we have not specified the bundles whose connections one is to integrate
over. Just as for the path integral on $\Sigma_{g} \times S^{1}$, upon
diagonalisation, one must sum over those non-trivial $U(1)$-bundles that
arise as obstructions to diagonalisation. As we have argued, these are
precisely the torsion bundles on $M_{(g,p)}$. We deal with the question
of how to implement this concretely and other related issues in the
next section.

Before turning to these questions we wish to compare what we have done
here with the calculations for $p=0$ in \cite{btg/g}. In reducing
Chern-Simons theory on $\Sigma_g \times S^1$ to an Abelian theory
on $\Sigma_g$, one can either diagonalise first and then reduce to 2
dimensions (this is also the strategy that we have adopted here) or one
can reduce first to a non-Abelian 2-dimensional theory (the $G/G$ model)
and then apply Abelianisation.  In \cite{btg/g} we perfomed the detailed
calculations of the determinants for the latter approach. One finds that,
when the determinants arising from the (chiral, 2-dimensional) $G/G$-model
are $\zeta$-function regularised, they give not only the Ray-Singer torsion
(arising in this approach from the Weyl integral formula) but also the
phase shift in the level, $k \rightarrow k+ c_{\lg}$, together with some
normalisation terms. Had we done the calculation the other way around,
starting from the Abelian (and less chiral) 3-dimensional theory, then
we would have found that the shift in the level arises not from the
$\zeta$-function but from the $\eta$-function regularisation of the phase
of the determinant acccompanying the Ray-Singer torsion. Indeed the
calculations of Appendix \ref{regs} are valid for $p=0$ as well and thus
complete this alternative calculation, only sketched in \cite{btg/g}.
A similar calculation has also been carried out in \cite{BW}.

It should be of interest to find an analogue of the second procedure
(reduce first and then diagonalise) also for $p\neq 0$, as this would
give a non-abelianised description of q-deformed Yang-Mills theory,
i.e.\ a $\phi F \ra \phi F + \phi^2$-like deformation of the $G/G$-model,
perhaps the Lagrangian realisation proposed in \cite{klimcik}.

\section{The Resulting Abelian Theory}\label{trat}

The Abelian curvature 2-form $F_{H}$ in (\ref{ssigma}) 
involves not only $A_{H}^{\lt}$ but also a component of the
connection in the $\kappa$ direction (the curvature is nevertheless 
horizontal). Indeed, we know that when we diagonalise, non-trivial
$T$-bundles arise; and  from the discussion in Appendix \ref{apgysin}
we know that these are torsion bundles. 

Consider $G=SU(2)$ and $T=U(1)$: a line
bundle $L$ on $\Sigma_{g}$ has first Chern class $c_{1}(L) = r [\omega]$,
so that $\pi^{*}(L)$ has first Chern class 
\be
c_{1}(\pi^{*}(L)) = r \, [\pi^{*}(\omega)]= \frac{r}{p}\, [d\kappa]\;\;.
\ee
We thus see that the pull-back connection may be taken to be
\be
\mathsf{A}= 2\pi \frac{r}{p} \kappa\;\;,
\ee
which, as announced, lives in the $\kappa$-direction.
It is perhaps somewhat surprising that, even though we have split off the part
of the $G$-connection in the direction of $\kappa$,  we are forced to
reintroduce an (albeit non-dynamical) 
component in that direction upon diagonalisation.

This connection has holonomy in the $S^{1}$ direction of $M_{(g,p)}$,
\be
\exp{( i \oint \mathsf{A})} 
= \exp{( 2\pi i \frac{r}{p}) } \in \mathbb{Z}_{p}\label{hol}
\ee
and captures the torsion.  The curvature 2-form $F_H$ appearing in
(\ref{sm}, \ref{ssigma}) is then
\be
F_{H} = dA_{H} + d\mathsf{A} = dA_{H} + 2\pi \, \frac{r}{p}\, d\kappa 
\label{curve}
\ee
and the path integral should include a summation over $r=0,\ldots,p-1$.

This argument generalises to higher rank. Normalising the component
fields by expanding
$\phi$ and $A_H$ in a basis of simple roots,
\be
\phi = \sum_{i=1}^{\rk} \phi^{i} \, \alpha_{i}, \;\;\; A_{H} =
\sum_{i=1}^{\rk} A^{i}\, \alpha_{i}\;\;,
\ee
$F_H$ in (\ref{2dpi}) has the form
\be
F_{H} = \sum_{i=1}^{\rk} \left( dA_{H}^{i} + 2\pi \, \frac{r^{i}}{p}\,
  d\kappa \right) \, \alpha_{i}\;\;.
\label{2dfh}
\ee

\subs{Suming Over Bundles and a Symmetry}

Our task is to sum over all allowed torus bundles, that is all torus
bundles of finite order, on $M$. We should therefore sum over all
allowed values of $r^{i}= 0, \dots, p-1$ of (\ref{2dfh}). But how does
the path integral (\ref{2dpi}) know that $r^{i}=0$ is the same as $r^{i}=p$?

Note that shifting the $r^i$ by multiples of $p$, $r^i \ra r^i + p\gamma^i$, 
$\gamma^i\in\ZZ$ is tantamount to shifting $F_H$ by an element
$2\pi\gamma = 2\pi\gamma^i\alpha_i$ of the integral lattice
$I=2\pi\ZZ[\alpha_i]$ of $G$. Thus consider
the transformation
\be
F_{H} \rightarrow F_{H} + 2\pi \, d\kappa\, \gamma \;\;\;\;\;\;
 \phi \rightarrow
\phi - 2\pi \gamma \;\;.\label{shifts}
\ee
We claim that this is an invariance of the path integral (\ref{2dpi}).
Indeed, even though the exponent is not invariant, it changes by
\be
-i \pi (k+c_{\lg}) p\, \sum C_{mn}\gamma^{n}\gamma^{n}\;\;.\label{phaset}
\ee
Here $C_{mn}$ is the Cartan matrix
\be
C_{mn}= \Tr \alpha_{m}\, \alpha_{n}\;\;.
\ee
Now $C_{mn}$ is a symmetric integral matrix with even diagonal
entries and consequently $\sum C_{mn} \gamma^{m}\gamma^{n}$
is an even integer. Thus the phase
(\ref{phaset}) is $2\pi i t$ for some integer $t$
and the exponential is invariant.

The Ray-Singer torsion term is also invariant under these transformations
so they represent a symmetry of the theory at hand. This is consistent
with the fact that we should only sum over the torsion classes,
the symmetry guaranteeing that the result does not depend on the
representative.

It is worth noting that, for $p\neq 0$,
the symmetry (\ref{shifts}) is not a symmmetry of
the original theory.  Rather it reflects an ambiguity in our description
of functions on $M$ as sections of line bundles on $\Sigma_{g}$. Indeed,
as we already discussed in section 2, the pull-back of the $p$-th
power of any line bundle on $\Sigma_g$ to $M_{(g,p)}$ is trivial.
Hence upon pull-back sections of a line bundle $L$ on $\Sigma_g$ are
indistinguishable from sections of $L \otimes \mathcal{L}^p$ for some
line bundle $\mathcal{L}$. This is the origin of the ambiguity, which
thus consistently appears as a symmetry of the theory.

This should be contrasted with what happens for $p=0$: in that case 
$d\kappa=0$ and the background connection $A$ has no $\kappa$-component.
Nevertheless 
the symmetry $\phi \rightarrow \phi + 2\pi \gamma$ exists as it is
part of the original gauge symmetry (large Abelian gauge transformations
wrapping around the $S^1$). This gauge symmetry leads to $\phi$ being a
compact scalar taking values in $\lt/I$. While compactness of $\phi$ is not
required by the gauge symmetries for $p\neq 0$, 
we will see shortly that the symmetry (\ref{shifts}) can nevertheless
be used to provide an alternative description of the same theory in terms
of a compact scalar (thus establishing the equivalence with the q-deformed
Yang-Mills model of \cite{AOSV}).

\subs{Reduction to Finite Dimensional Integrals}

We had already argued at the end of section \ref{CSonM} that ultimately only
constant $\phi$ contribute to the path integral. To see this more
explicitly, consider the $\int \phi dA_H^{\lt}$ part of the action 
(\ref{ssigma}). The integral over
$A_{H}^{\lt}$ imposes the required condition 
\be
d\phi^{i} =0\;\;.
\ee
More precisely, as 
we have already explained, the harmonic parts of $A_{H}^{\lt}$
do not enter in the action but just lead to an overall normalisation
of the path integral. The exact parts of the gauge field do not appear
either, but these are the components which are gauge degrees of
freedom (the residual $U(1)^{\mathbf{rk}}$ gauge symmetry). That only leaves
the co-exact piece of the gauge field and integration over these
components imposes that the $\phi^{i}$ are constant.

With $\phi$ constant, the partition function (\ref{2dpi}) reduces to the
finite-dimensional integral
\be
Z_k[M_{(g,p)}, G] \sim 
\ex{4\pi i p\Phi_{0}}
\sum_{r \in \mathbb{Z}_{p}^{
    \mathbf{rk}}} \,  \int_{\lt} 
T_{S^{1}}(\phi)^{\chi(\Sigma_{g})/2}\, \exp{i \frac{k+
      c_{\lg}}{4\pi} \Tr \left( p \, \phi^{2} + 4 \pi r\, \phi
  \right)} \label{xfinal1}
\ee
In particular, this formula reproduces (\ref{rl}) directly (for a
judicious choice of normalisation) when we restrict to the Lens spaces
$L(p,1)$, take the group to be $G=SU(2)$, and deform the integration
contour appropriately (there are no residues to worry about).

At this point we also refer back to the discussion in the Introduction
where we compared and contrasted the general structure (\ref{bw}) of the
formulae obtained from non-Abelian localisation with the structure
(\ref{bt}) we found above that follows from diagonalisation and
integrating out all the 2-dimensional gauge fields.

\section{Compact versus Non-Compact Yang-Mills and Interpretations}

In this section we will look at what the solutions of these theories
implies from the physics and mathematics points of view. Our starting
point is that we can sum over all line bundles and declare the shift
symmetry (\ref{shifts}) to be a symmetry of the resulting Abelian theory
(even though it is not a symmetry of the original theory).

Then given that we have this symmetry we should gauge fix it. One way
to do that is to declare that we only sum over the required range of
$r$. An alternative gauge fixing would be to compactify $\phi$ as is
done in \cite{AOSV}. Surprisingly enough the `parent' theory (that is
the one still to be gauge fixed) is just ordinary non-compact Yang-Mills
theory on $\Sigma_{g}$ with a gauge invariant observable inserted!

We first show directly that the two gauge fixed partition functions agree so
that they indeed arise from one `parent' partition function. For any
function $f(\phi)$ on $\lt$, invariant under shifts in the integral lattice
$I$, and any integer $q$ (for us $q = (k + c_{\lg})$), let
\be
Z_{q,p}(f) = \sum_{r \in \mathbb{Z}_{p}^{\mathbf{rk}} }\int_{\lt}
\, f(\phi)\, \exp{ i\frac{q}{4\pi} 
  \Tr\left( p \phi^{2} + 4\pi r\phi \right)}\label{cnc}
\ee
be a model of the partition function (\ref{xfinal1}), 
and denote by $\tilde{Z}_{q,p}(f)$ the analogous partition function
for a compact scalar,
\be
\tilde{Z}_{q,p}(f) = \sum_{r \in \mathbb{Z}^{\mathbf{rk}}}\int_{\lt / I} \,
f(\phi) \,  \exp
  i\frac{q}{4\pi}\Tr \left( p \phi^{2} + 4r\pi \phi \right) \label{cnc2}
\ee
We now establish the identity
\be
Z_{q,p}(f) = \tilde{Z}_{q,p}(f) \;\;.
\label{cnc3}
\ee
To that end we note that the integral over
$\lt$ is the same as integrating each lattice cell and then summing
over the cells. However, one can shift any cell to a given one by an
element of $I$. The integral over the given cell is then an integral
over $T = \lt / I$. $f(\phi)$ is, by hypothesis, invariant under such
shifts. The exponent in (\ref{cnc}), however, is not. Taking into
account the shift one has
\be
Z_{q,p}(f) = \sum_{s \in \mathbb{Z}_{p}^{\mathbf{rk}} }\sum_{n \in
  \mathbb{Z}^{\mathbf{rk}}} \int_{\lt / I 
} \, f(\phi) \, \exp{ 
  i\frac{q}{4\pi}\Tr \left( p \phi^{2} + 4(s+ pn)\pi \phi \right) }
\ee
As one sums over $n$ and $s$ the combination
\be
r = np + s \label{expans}
\ee
with $n \in \mathbb{Z}^{\mathbf{rk}}$ and $s \in
\mathbb{Z}_{p}^{\mathbf{rk}}$, covers $\mathbb{Z}^{\mathbf{rk}}$.
This establishes (\ref{cnc3}).

In particular, therefore, an equivalent way of writing the partition
function (\ref{xfinal1}) is
\be
Z_k[M_{(g,p)}, G] \sim \ex{4\pi i
    p\Phi_{0}} \sum_{r \in \mathbb{Z}^{
    \mathbf{rk}}} \,  \int_{\lt/I} 
T_{S^{1}}(\phi)^{\chi(\Sigma_{g})/2}\, \exp{i \frac{k+
      c_{\lg}}{4\pi} \Tr \left( p \, \phi^{2} + 4 \pi r\, \phi
  \right)} \label{xfinal2}
\ee
This formula has been obtained at the level of the constant modes. But
it shows us that the field theory which had a non-compact scalar
$\phi(x)$ and a finite sum over torsion bundles is equal to the field
theory written down in \cite{AOSV}, with a compact scalar and a 
$\phi^2$-term. Written this way the 
theory only makes sense with the infinite sum over bundles.

Now we turn to the theory which is to be gauge fixed to give either of
the equivalent partition functions above. The parent theory is simply
\be
W_{q,p}(f) = V^{-1}\sum_{r \in \mathbb{Z}^{\mathbf{rk}} }\int_{\lt}
\, f(\phi)\, \exp{ i\frac{q}{4\pi} 
  \Tr\left( p \phi^{2} + 4\pi r\phi \right)}\label{parent}
\ee
where $V=\mathbf{Vol}(p\mathbb{Z}^{\mathbf{rk}})$ is the `gauge volume'. 
Fixing the gauge by passing from $r\in  \mathbb{Z}^{\mathbf{rk}}$ to
$r\in \mathbb{Z}_p^{\mathbf{rk}}$ one recovers $\tilde{Z}_{q,p}(f)$.
Alternatively, writing
$r$ once more as in (\ref{expans}) and shifting $\phi$ by the
integral lattice elements to eliminate the $n$-dependence in the
action one finds $Z_{q,p}(f)$.

\subs{How Compact is Compact Yang-Mills?}

Apart from compactness one of the main difference between the abelianised
Chern-Simons theory on $M_{(g,p)}$, when finally reduced to a theory on
$\Sigma_{g}$, and abelianised Yang-Mills theory on $\Sigma_{g}$ is that
the measure for the former involves $\det{(1 - \ex{\ad \phi})}$ (related
to the Weyl integral formula for Lie groups) and $\det{(\ad\phi)}$
(arising from the Weyl integral formula for Lie algebras) for the latter,
both raised to the power $\Sigma_{g}/2$. 

These two measures are related by the Jacobian of the exponential map
on Lie groups, given (upon restriction to $\lt$) by the function
\be
j_{\lg}(\phi) = \frac{\det_{\lk}{(1 - \ex{\ad \phi})}}{\det_{\lk}{(\ad\phi)}}
\ee
which makes an apperance in the study of coadjoint orbits and equivariant
localisation - see e.g.\ \cite{BGV}.

Another difference rests in the normalisation of the action. Following
the $\phi F_{A}$ terms around we see that there is an extra factor of
$q$ in front of the Chern-Simons action compared to the Yang-Mills
action in the normalisation adopted in \cite{btictp}. This can be
compensated for in the Yang-Mills theory if we send $\phi \rightarrow
q \phi$. With this change understood, if we consider the expectation value of 
\be
j_{\lg}(\phi)^{\chi(\Sigma_{g})/2}\label{jchi}
\ee
in Yang-Mills theory on $\Sigma_{g}$, with a very particular value for
the area, then on Abelianisation \cite{btictp} we would arrive at
(\ref{parent}) 
with $f(\phi)$ given by the Ray-Singer torsion raised to half the
Euler characteristic. Hence one is, up to possibly some
normalisation factors, evaluating the Chern-Simons
path integral on $M_{(g,p)}$.

What we have just exhibited is that the partition function and gauge
invariant observables in compact Yang-Mills can be equally thought of
as arising from non-compact Yang Mills theory but with the insertion
of the operator (\ref{jchi}).

\subs{Cohomology of Yang-Mills Connections on $\Sigma_{g}$}

Witten \cite{ewym} has shown how to pass from the cohomological
Yang-Mills theory on $\Sigma_{g}$ to the physical theory. In this way
pairings of the cohomology ring on the moduli space of flat
connections on $\Sigma_{g}$ can be extracted from the physical partition
function. Higher critical points of the Yang-Mills action contribute
to the path integral as well but these are exponentially supressed as
the area goes to zero.

In the present situation we cannot let the area go to zero since it is
linearly related to $p$. Hence the contribution to the path integral,
from the point of view of non-Abelian localisation \cite{ewym,BW},
includes components of the moduli space of Yang-Mills connections
which are not flat.

We have already established that Chern-Simons theory on $M_{(g,p)}$ is
equivalent to Yang-Mills theory on $\Sigma_{g}$ for a particular area.
Consequently we now have that the Chern-Simons partition function
yields certain intersection pairings on the moduli space of Yang-Mills
connections on $\Sigma_{g}$. One can obtain precise formulae relating
the two as has been done in \cite{BW}. 

\subs{The Weyl Group}

The formulae above are invariant under the action of the
Weyl group $W$ which is part of the original gauge group. We thus need
to divide by the ``volume'' (or mod out by the action) of $W$.

For $SU(2)$ the Weyl group
is $\mathbb{Z}_{2}$  and the
non-trivial element sends the connection $A$ to $-A$. As we emphasised
previously it is as if we are working with a connection that is of the
form
\be
A= A_{H} + \phi \, \kappa + 2\pi \frac{r}{p} \kappa
\ee
and given the range of $r$ in (\ref{xfinal1}) the Weyl transformation
corresponds to
\be
A_{H} \rightarrow - A_{H}, \;\;\; 
\phi \rightarrow -\phi - 2\pi , \;\;\; r\rightarrow p-r. \label{weyl}
\ee 
When we abelianise on
$\Sigma_{g}$ we are essentially splitting the $SU(2)$ bundle in a
direct sum of lines 
\be
\ad_{\mathbb{C}} (P_{SU(2)}) = \mathcal{O}(-2n) \oplus
\mathcal{O} \oplus \mathcal{O}(2n)
\ee
and the Weyl group acts to exchange $\mathcal{O}(-2n)$ and
$\mathcal{O}(2n) $. The Chern classes $n$ and $-n$ do not necessarily
agree $\mathrm{mod}\, p$ and so do not correspond to the same torsion
class on $M_{(g,p)}$. Rather $-n$ for $n\geq 0$ is the same as $p-n$ and
explains the last transformation in (\ref{weyl}). 

On the other hand, once one has transformed to (\ref{xfinal2}) one sees
that the Weyl group acts by $\phi \rightarrow - \phi$ and $r
\rightarrow -r$. Indeed if one keeps track of the field redefinitions
then these transformations agree with (\ref{weyl}). The formula that
we will use mostly in the following is
\be
Z_k[M_{(g,p)}, G] = \Lambda\,\ex{4\pi i p\Phi_{0}}\,  \sum_{r \in \mathbb{Z}^{
    \mathbf{rk}}} \,  \int_{\lt/\Gamma^{W}} 
T_{S^{1}}(\phi)^{\chi(\Sigma_{g})/2}\, \exp{i \frac{k+
      c_{\lg}}{4\pi} \Tr \left( p \, \phi^{2} + 4 \pi r\, \phi
  \right)} \label{xfinal2a}
\ee
where $\Gamma^{W} = I \rtimes W$ is the affine Weyl group and $\Lambda$ is
a real normalisation constant that remains to be determined.

Notice that the Ray-Singer torsion has zeros at the boundary of the Weyl
chamber, which means that for genus $g>1$ the integrals (\ref{xfinal1},
\ref{xfinal2a}) diverge. As already indicated in \cite{btg/g} the
way around this is to regularise by giving a small mass term to
the connection, while preserving the residual $U(1)^{\mathbf{rk}}$
invariance. It is easier to incorporate this in the form (\ref{xfinal2a})
since this amounts to not including the boundaries of the Weyl chamber. As
we will see presently the contributions to the path integral are at
discrete points and so the regularisation renders the integrals finite.

\section{The Partition Function on $M_{(g,p)}$}\label{partfm}

We have thus found the following equivalent finite-dimensional intgegral
expressions for the partition function on $M_{(g,p)}$,
\bea
\hspace{-.5cm}Z_k[M_{(g,p)}, G] &\!=\!& \Lambda\frac{\ex{4\pi i p\Phi_{0}}}{|W|}
\sum_{r \in \mathbb{Z}_{p}^{
    \mathbf{rk}}}   \int_{\lt} 
T_{S^{1}}(\phi)^{\chi(\Sigma_{g})/2} \exp{i \frac{k+
      c_{\lg}}{4\pi} \Tr \left( p  \phi^{2} + 4 \pi r \phi
  \right)} \label{final1}\\
 &\!=\!& \Lambda \frac{\ex{4\pi i
    p\Phi_{0}}} {|W|} \sum_{r \in \mathbb{Z}^{
    \mathbf{rk}}}   \int_{\lt/I} 
T_{S^{1}}(\phi)^{\chi(\Sigma_{g})/2} \exp{i \frac{k+
      c_{\lg}}{4\pi} \Tr \left( p  \phi^{2} + 4 \pi r \phi
  \right)} \label{final2}\\
 &\!=\!& \Lambda\ex{4\pi i p\Phi_{0}}  \sum_{r \in \mathbb{Z}^{
    \mathbf{rk}}}   \int_{\lt/\Gamma^{W}} 
T_{S^{1}}(\phi)^{\chi(\Sigma_{g})/2} \exp{i \frac{k+
      c_{\lg}}{4\pi} \Tr \left( p  \phi^{2} + 4 \pi r \phi
  \right)} \label{final2a}
\eea
with $\Lambda$ real.

One could have kept tabs on most of the normalisation factors.
It is more convenient, however, to give a prescription for fixing
$\Lambda$. One takes $\Lambda$ to be independent of $p$ and chosen
so that for $p=0$ (\ref{final2a}) agrees with the partition function
obtained on $\Sigma_{g} \times S^{1}$ in \cite{btg/g}, that is, it
reproduces the Verlinde formula for dimension of the space of conformal
blocks on $\Sigma_{g}$.\footnote{For $G=SU(2)$ this fixes $\Lambda =
\left(2(k+2)\right)^{g}/2\pi $.}

Once one has fixed the normalisation as we have then the partition
function on $M_{(g,p)}$ is the same as the expectation value of an
operator in the theory on $\Sigma_{g} \times S^{1}$,
\be
Z_k[M_{(g,p)}, G ] = \ex{4\pi i p\Phi_{0}}\, \langle
\ex{ip\frac{(k+c_{\lg})}{4\pi} \Tr 
  \phi^{2} }
\rangle_{\Sigma_{g} \times S^{1}} 
\ee
where on the right hand side $\phi$ is the constant mode of the gauge
field in the $S^{1}$ direction and we have already imposed that $\phi$ is in
the Cartan sub-algebra (and, as discussed before, $\phi$ is compact in 
the G/G-model). The difficulty in finding the correct non-gauge fixed
version of the operator $\exp{i\frac{(k+c_{\lg})}{4\pi} \Tr \phi^{2} }$
is at the heart of the problem of developing a $G/G$ type model
corresponding to $M_{(g,p)}$.

Nevertheless, this is a remarkable (and remarkably simple) result. It
states that manifolds of non-trivial Chern classes are simply created by
insertions of the operator $\exp{i\frac{(k+c_{\lg})}{4\pi} \Tr \phi^{2}
}$ in the path integral for the trivial bundle. That this should be
so has first been argued by Vafa \cite{vafaym} in a rather different
way. Here we have provided a path integral derivation of this fact.

As in \cite{btg/g}, the sum over $r \in \mathbb{Z}^{\mathbf{rk}}$
imposes a $\d$-function constraint
on the field $\phi$ to configurations that satisfy
\be
\phi = \frac{2\pi (\lambda + \rho)}{k+ c_{\lg}}\label{phiconst}
\ee
for some weight $\lambda$. Here $\rho$ is the Weyl vector, one half the
sum of the positive roots. In particular we have
\be
\frac{k+c_{\lg}}{4\pi}\Tr \phi^{2} =  \frac{\pi}{(k+ c_{\lg})}
\Tr((\lambda + \rho)^2 
=- \frac{2\pi}{(k+
  c_{\lg})}\left( c_{2}(\lambda) + c_{\lg}\frac{\dim G}{24}\right)
\ee
where the quadratic Casimir is
\be
c_{2}(\lambda) = -\frac{1}{2}\Tr\left(\lambda^2 + 2\rho\lambda\right)
\ee
and we have made use of the Freudenthal - de Vriess formula 
\be
\Tr \rho^2 = - c_{\lg} \frac{\dim G}{12}\;\;.
\ee
Combining this with the overall phase we have that
\be
\frac{k+c_{\lg}}{4\pi}\Tr \phi^{2} + 4\pi \Phi_{0} = - 2\pi
\Phi(\lambda) 
\ee
where
\be
\Phi(\lambda) =
\frac{1}{(k+ c_{\lg})} \left( 
c_{2}(\lambda) - \dim G \frac{k}{24} \right).
\ee
Hence the $T$ matrix
\be
T_{\lambda \mu} = \d_{\lambda\mu} T_{\lambda}\;\;\;\;\;\;\;\;
T_{\lambda} = \exp{2\pi i \Phi(\lambda)}\label{T}
\ee
appears naturally in the evaluation of the path integral.

The compactness of $\phi$, $\phi\in \lt/\Gamma^{W}$,
constrains the possible weights so that they are integrable highest
weights at level $k$. 
Writing the partition function on $\Sigma_{g} \times S^{1}$ as
\be
Z_k[\Sigma_{g} \times S^{1}, G] = \sum_{\lambda} S_{0\lambda}^{2-2g}
\ee
where the sum is over integrable highest weights, with our normalisation
one then has
\be
Z_k[M_{(g,p)}, G] = \sum_{\lambda} S_{0\lambda}^{2-2g} \,
T_{\lambda}^{-p} \;\;.\label{stpi}
\ee
Note that both the $T$ and $S$ matrices are the standard modular
matrices for the Wess-Zumino-Witten model. $T$ is diagonal (in the
Verlinde basis) and $S$ is symmetric \cite{WCS}. These mapping class
group generators are subject to the relations
\be
S^{2}=
\mathbb{I}, \;\;\; (ST)^{3} = \mathbb{I} \;\;.\label{maprels}\;\;.
\ee

We wish to interpret the results of performing the path integral
in terms of the surgery prescription of Chern-Simons theory. In order
to do that we need to introduce Wilson loops which is what we do next.

\section{Surgery and the Framing Provided by the Path Integral}\label{wl}

Since (\ref{stpi}) is already reminiscent of a surgery formula we would
expect that indeed
\be
Z_{k}[M_{(g,p)}, G] = \sum_{\lambda} K_{0\lambda}^{(p)} \,
Z_{k}[\Sigma_{g}\times S^{1}, G, R_{\lambda}]
\ee
for some mapping class group element $K^{(p)}$. In this and subsequent
formulae $R_{\mu}$ is a Wilson loop in the vertical direction in the
representation $\mu$. Since it is the integrals over
$A_{H}^{\lt}$ which ensure that $\phi$ is constant we see that the inclusion
of Wilson lines into the path integral does not change
this. So we have not indicated any base point dependence
as at the end of the day there is no such dependence. We know from
$\cite{btg/g}$ that 
\be
Z_{k}[\Sigma_{g}\times S^{1}, G, R_{\lambda}] = \sum_{\sigma}
S_{0\sigma}^{1-2g}\, S_{\lambda \sigma}
\ee
and so comparing with (\ref{stpi}) suggests that
\be
K^{(p)} = ST^{-p}S \label{kp}
\ee
which allows us to write $K^{(p)}= K^{p}$ where
\be
K= ST^{-1}S\;\;.\label{K2}
\ee

The canonical choice for obtaining $S^{3}$ from a surgery on
$S^{2}\times S^{1}$ 
is to identify 
up to the action of $S$, however, quite generally the surgery
\be
K = T^{m}ST^{n} \label{S3surg}
\ee
yields $S^{3}$ with $m$ and $n$
arbitrary integers. Different values of $m$ and $n$ correspond to
different framings of $S^{3}$. The path integral calculation will correspond to
one of the possible matrices in (\ref{S3surg}), that is it will
correspond to a definite value for $m$ and $n$ and hence to a
particular framing of $S^{3}$. 

Witten \cite{WCS} has shown that in the
canonical framing one has $K=S$.\footnote{Note that with this choice of $K$
one cannot possibly have $K^{(p)}=K^p$ since $S^2=1$.}
By making use of the second relation in (\ref{maprels}) we see that we can
write $K$ as
\be
K = TST \label{K1}
\ee
suggesting that $m=n=1$ so that we are not working in the
canonical framing if our identification (\ref{kp}) is correct.

As an aside we note that one could have used the non-compact
representation (\ref{final1}) of the partition function rather than
(\ref{final2a}) to calculate the Hopf link. We show in Appendix C
that for $G=SU(2)$ this leads directly to $K=TST$ without invoking
(\ref{maprels}). Since we have calculated the same quantity $K$
in two different ways, we thus have a path integral derivation of
the relation $TST = ST^{-1}S$.

Consider now the more general surgery formulae involving two vertical
Wilson lines
\be
Z_{k}[M_{(g,p)}, G, R_{\mu}\, R_{\nu}] = \sum_{\lambda} K_{\mu\lambda}^{(p)} \,
Z_{k}[\Sigma_{g}\times S^{1}, G, R_{\lambda}\, R_{\nu}] \label{2ndsurg}
\ee
(the vertical Wilson line $R_{\nu}$ is not in the tubular
neighbourhood where the surgery takes place but $R_{\mu}$
is). In particular for $g=0$, $p=1$ one is calculating the expectation
value of the Hopf link since, on noting that $Z_{k}[S^{2}\times
S^{1}, G, 
R_{\lambda}\, R_{\nu}] = \delta_{\lambda \nu}$, one finds
\be
K_{\mu \nu} = Z_{k}[S^{3}, G, R_{\mu}\, R_{\nu}] \;\;.\label{kpi}
\ee
The left hand side of (\ref{2ndsurg}) is easy to
determine given the constraint (\ref{phiconst}) implied in (\ref{final2a}),
\be
Z_{k}[M_{(g,p)}, G, R_{\mu}\, R_{\nu}] = \sum_{\lambda}
S_{0\lambda}^{-2g} S_{\lambda \mu} S_{\lambda \nu} T^{-p}_{\lambda}
\;\;.\label{2wl} 
\ee

Comparing with (\ref{2ndsurg}) we see that indeed $K^{(p)}$ is as in 
(\ref{kp}) and that $K$ is given by (\ref{K2}) and consequently
(\ref{K1}). Indeed one obtains the formula for $K$ directly by setting
$g=0$, $p=1$ in (\ref{2wl}).

We read off from (\ref{kpi}) and (\ref{K1}) that we have framed the
vertical knots by +1 units from the canonical framing of the knot. This
is related to the -2 unit of framing for $S^{3}$ that we are away from
the canonical framing as determined by Beasley and Witten \cite{BW}. In
the current setting we see this framing of $S^{3}$ on setting $\mu=\nu=0$.

Beasley and Witten \cite{BW} determine the change of framing in a related
but somewhat different manner. The advantage of our approach is that it
allowed us to identify the element, $K^{(p)}$, of the mapping class group
with which one builds non-trivial bundles and which, at the same time,
incorporates the change of framing.

The operator $K=TST$ was also identified in \cite{akmv} as the operator
that generates the insertion (\ref{insertion}) of the Chern class in 
the matrix model.
On the other hand, comparing with \cite{deharo2}, we note that 
q-deformed Yang-Mills theory,
as defined there, differs from our definition by the (insignificant) change
$p \ra -p$, accounting for the difference 
\be
K=ST^{-1}S \ra K^\prime = STS = T^{-1}ST^{-1}\;\;.
\ee
Indeed, if one uses $K^\prime$ instead of $K$ in the surgery prescription, 
one finds that the partition function is given by (\ref{stpi}) with $p \ra
-p$.

\section{Generalisations}

There are various generalisations that are possible. These include the
choice of other gauge groups. We have not considered either
non-compact gauge groups or compact but not simply connected gauge
groups thus far in this paper. Also one may ask about other
3-manifolds, do our considerations above apply?

The steps involved in solving the theory on $U(1)$ bundles were:
\begin{enumerate}
\item Decomposing the gauge field as (\ref{decomp}).
\item Fixing the gauge (\ref{gf}).
\item Diagonalising (\ref{gfc})
\item Pushing the calculations down to the base $\Sigma_{g}$.
\end{enumerate}

Step 1 is always possible since one can introduce a contact structure
on any 3-manifold and this automatically gives us the required
decomposition. However, $\kappa$ may not be globally defined in which
case $\phi$ is not globally defined either but is correctly thought of as a
section of some real line bundle. $\kappa \phi$ is globally well defined.

Step 2 now follows but one must ensure that the derivatives in the
gauge fixing condition are
covariant derivatives with respect to the real line bundle.

Step 3 is not straightforward in general. Compact but not simply
connected groups pose a problem. While we can
indeed abelianise gauge theories for 
non-simply connected groups on a Riemann surface (see \cite{btdiag} for a
general discussion and \cite{bhtkz} for concrete applications),
there are other obstructions to
diagonalisation in 3 dimensions and above \cite{btdiag}. 
Non-compact groups are
already an issue on Riemann surfaces. The problem here is that there
may be different maximal tori to which one conjugates. Likewise, there is a
problem in extending Abelianisation to other 3-manifolds again due to
obstructions beyond the ones we have already considered. However, there is the
hope that the technique will be applicable in the general Seifert
fibred case, since this amounts to establishing a diagonalisation for
2-dimensional orbifolds, but that is still to be shown.
 
Step 4 is certainly a technical convenience but may not be essential,
otherwise for a general 3-manifold we would not know how to proceed.

In the above discussion it is step 3 which is the most difficult to
mimic in other cases. However, for special non-compact groups one may
apply all of the machinery. We finish with an example of this.

3-dimensional BF-theories, topological gauge theories with action
\be
S_{BF} = \int_{M} \, \Tr\, B\wedge F_{A} \label{bf}\;\;,
\ee
where $B$ is a Lie algebra valued 1-form, can be interpreted as
Chern-Simons theories with a non-compact gauge group usually denoted
by $IG$ or $TG \sim G \times \lg$ \cite{WBF,tftpr}.
There is no sufficiently well understood surgery/CFT prescription for
calculating the partition function of pure BF theory and a direct path
integral evaluation of the partition function is desirable.

On 3-manifolds which have isolated flat connections, the BF path
integral is formally a sum of the Ray-Singer torsions of each of the
flat connections. When the flat connections are not isolated, there are
$B$ zero modes and one gets an integral over the tangent bundle of the
moduli space of flat connections. To avoid that we fix our attention
on the Lens spaces $L(p,1)$. One can now use the analysis of the previous
sections to get explicit formulae.

Apart from the usual gauge
symmetry, the action (\ref{bf})  has the invariance 
\be
B \rightarrow B + d_{A} f \label{bgt}
\ee
To perform calculations we decompose $A$ as before and $B$ as
\be
B= B_{H} + \kappa \, \lambda.
\ee
With this decomposition the action (\ref{bf}) becomes
\be
\int_{M}\Tr\, \left( B_{H} \wedge d_{\phi}A_{H} + B_{H}\wedge d(\kappa \,
 \phi) + \lambda \, \kappa \wedge d (\kappa \, \phi) + \lambda
\, \kappa \wedge d_{\phi} A_{H}  + \lambda \kappa \wedge A_{H}
\wedge A_{H}\right)
\ee
We impose the gauge conditions (\ref{gf}) and (\ref{gfc}) on $\phi$ and similar
conditions on $\lambda$,
\be
L_{K}\lambda = 0, \;\;\; \lambda^{\lk}=0.
\ee
Now the path integral over invariant $A_{H}$ implies that the $\lambda$
are constant and the integral over invariant $B_{H}$ implies that the
$\phi$ are constant. The other integrals together with the ghost terms
yield the Ray-Singer torsion, so (for $G=SU(2)$ for simplicity)
\be
Z_{BF}[L(p,1),G=SU(2)]= \sum_{r=0}^{p-1}\int d\lambda\, d\phi \;
T_{S^{1}}(\phi)^2 \, 
\exp{\left( ip
  \Tr \lambda \, 
  \phi + i r \, \lambda \right)}
\ee
The integral over $\lambda$ gives us a delta function on $\phi$ so
that $\exp\phi$ is a root of unity,
\be
Z_{BF}[L(p,1),SU(2)]=\sum_{r=0}^{p-1}T_{S^{1}}(2\pi i r/p)^2=
\sum_{r=0}^{p-1} T_{L(p,1)}(2\pi i r/p).
\ee
One can also evaluate the expectation values for Wilson loops in the
fibre direction with and without the $B$ field. We leave that for
another occassion.

Finally, we expect that super $BF$ theory can be solved in a similar way on
Lens spaces, thus giving us the $SU(n)$ Casson invariant in these cases.

\subsection*{Acknowledgements}

We are grateful to M.S.\ Narasimhan and T.R.\ Ramadas for Gysin-esque
discussions. This work has been partially
supported by the Swiss National Science
Foundation and by the EU under contract MRTN-CT-2004-005104.

GT would like to express his gratitude to Bob Delbourgo
for his support and friendship all these years. 
\appendix

\section{Diagonalisation and Torsion Bundles on $M_{(g,p)}$}\label{apgysin}

\subs{Line Bundles on $M_{(g,p)}$ and $\Sigma_g$ and the Gysin Sequence}

The 3-manifolds $M_{(g,p)}$ that we are interested in are principal 
$U(1)$-bundles over a 2-dimensional surface $\Sigma_{g}$ of genus $g$, 
$U(1) \rightarrow M_{(g,p)} \stackrel{\pi}{\rightarrow}\Sigma_g$. 
There is a first
Chern class associated with this fibration which we denote by $-p$.
Thinking of $M_{(g,p)}$ as an $SO(2)$ bundle this is the real Euler class, 
while
the associated complex line bundle on $\Sigma_g$ is $\mathcal{O}(-p)$.
We will need to know various basic things about line bundles on
$M_{(g,p)}$ and their relations to line bundles on $\Sigma_g$.

To fix notions a bit we need some background. Given a vector field $v$
on $\Sigma_{g}$ we can lift it to $M_{(g,p)}$ by demanding that its
lift, $\widehat{v}$, satisfies $\iota_{\widehat{v}}.\kappa =0$ (this
picks out horizontal subspaces). In terms of the local coordinates
(\ref{kap}) this means that $\widehat{v} = (v,- a.v)$. Also, in the
standard manner, $U(1)$-equivariant complex valued
functions $f_m$ on $M_{(g,p)}$, with our 
normalisation of the $U(1)$-generator $K$, satisfy
\be
L_K f_m = 2\pi i m f_m .
\ee
Covariant
derivatives along $v$ are just actions of the vector field
$\widehat{v}$ on equivariant functions on $M_{(g,p)}$,
\be
\widehat{v}.f_{m} = v^{i}\left(\partial_{i} - a_{i}
\frac{\partial}{\partial \theta} \right) \, f_{m} =
v^{i}\left(\partial_{i} -2\pi i m a_{i} \right) f_{m} .
\ee
Consequently the equivariant functions $f_{m}$ are in one-to-one
correspondence with sections of the corresponding 
associated line bundle, i.e.\ with sections of $\mathcal{O}(mp)$ given
that the connection $a$ is a connection on $\mathcal{O}(-p)$ (\ref{fcc}).

Moreover, the pull-back $\pi^*\mathcal{O}(p)$ from $\Sigma_g$ to
$M_{(g,p)}$ is (tautologically) trivial. Hence 
\be
\left(\pi^*\mathcal{O}(m)\right)^{\otimes p}
\approx 
\pi^*\mathcal{O}(mp)
\approx 
\left(\pi^*\mathcal{O}(p)\right)^{\otimes m}
\ee
is also trivial for any $m \in \ZZ$. It follows that
the $p$-th power of the pull-back 
to $M_{(g,p)}$ of any line bundle $L$ on $\Sigma_g$ is trivial,
\be
\left(\pi^*L\right)^{\otimes p} \approx M_{(g,p)}\times\CC\;\;.
\ee
Since line bundles on $M_{(g,,p)}$ are classified by
$\mathrm{H}^{2}(M_{(g,p)},\mathbb{Z})$, $\pi^*L$ must be such that
\be
p c_1(\pi^*L) = 0 \in \mathrm{H}^{2}(M_{(g,p)},\mathbb{Z})\;\;,
\ee
i.e.\ that such line bundles correspond to $p$-torsion. 

We will now determine $\mathrm{H}^{2}(M_{(g,p)}, \mathbb{Z})$, show that it
indeed contains torsion $\mathbb{Z}_{p}$, and that all such torsion 
bundles arise via pull-back from $\Sigma_g$. We will address these issues 
using the Gysin sequence for sphere bundles.

Given a sphere bundle $M$ with fibre $F=S^{m}$
over a manifold $B$ one has the long exact sequence,
\be
\cdots\rightarrow \mathrm{H}^{n}(M) \stackrel{\pi_{*}}{\rightarrow}
\mathrm{H}^{n-m}(B) \stackrel{\wedge \, e}{\rightarrow}
\mathrm{H}^{n+1}(B)\stackrel{\pi^{*}}{\rightarrow}
\mathrm{H}^{n+1}(M)\rightarrow  \cdots
\ee
where $\pi_{*}$ is integration along the fibre (push down),
$\wedge e$ is wedging with respect to the Euler class and $\pi^{*}$ is
the usual pull back. The coefficients can be the integers.

In our case $m=1$, $F=S^{1}$, $B = \Sigma_{g}$ and the Euler class is
the first Chern 
class of the $U(1)$ bundle over $\Sigma_{g}$. We have, therefore,
\bea
0 \rightarrow H^{1}(\Sigma_{g}, \mathbb{Z})
\stackrel{\pi^{*}}{\rightarrow}  \mathrm{H}^{1}(M, \mathbb{Z})
\stackrel{\pi_{*}}{\rightarrow} 
\mathrm{H}^{0}(\Sigma_{g}, \mathbb{Z}) & & \nonumber \\ 
\stackrel{\wedge \, c_{1}}{\rightarrow}
\mathrm{H}^{2}(\Sigma_{g},
\mathbb{Z})&\noindent\stackrel{\pi^{*}}{\rightarrow}& 
\mathrm{H}^{2}(M, \mathbb{Z})\stackrel{\pi_{*}}{\rightarrow}
\mathrm{H}^{1}(\Sigma_{g}, \mathbb{Z}) \rightarrow 0. \label{long}
\eea

We see from the sequence that if $c_{1}$ is not trivial, as is the
case for the manifolds under consideration, then
\be
\mathrm{H}^{2}(\Sigma_{g}, \mathbb{Z})\stackrel{\pi^{*}}{\rightarrow}
\mathrm{H}^{2}(M, \mathbb{Z})
\ee
is surjective only when $\Sigma_{g}$ is the 2-sphere $S^{2}$,
i.e.\ only in that case do all line bundles on $M_{(g,p)}$ arise
from line bundles on $\Sigma_g$ via pull-back.

One can also read off, from the sequence, that while
$\mathrm{H}^{0}(\Sigma_{g}, \mathbb{Z})= \mathbb{Z}$ and
$\mathrm{H}^{2}(\Sigma_{g}, \mathbb{Z})= \mathbb{Z}$ are isomorphic,
that the map between the two is via multiplication by the integer
$p = c_{1}[\Sigma_{g}]$. As long as $p \neq 0$ this map has $0$ kernel
and so the image of $\mathrm{H}^{1}(M, \mathbb{Z})$ under $\pi_{*}$ is
$0$, so that we deduce that
\be
\mathrm{H}^{1}(M, \mathbb{Z}) \equiv H^{1}(\Sigma_{g}, \mathbb{Z}).
\ee
We can, therefore, re-write the sequence as
\be
0 \rightarrow 
\mathrm{H}^{0}(\Sigma_{g}, \mathbb{Z})
\stackrel{\wedge \, c_{1}}{\rightarrow}
\mathrm{H}^{2}(\Sigma_{g},
\mathbb{Z})\noindent\stackrel{\pi^{*}}{\rightarrow}
\mathrm{H}^{2}(M, \mathbb{Z})\stackrel{\pi_{*}}{\rightarrow}
\mathrm{H}^{1}(\Sigma_{g}, \mathbb{Z}) \rightarrow 0 \label{long2}
\ee
or
\be
0 \rightarrow \ZZ_p \stackrel{\pi^{*}}{\rightarrow}
\mathrm{H}^{2}(M, \mathbb{Z})\stackrel{\pi_{*}}{\rightarrow}
\mathrm{H}^{1}(\Sigma_{g}, \mathbb{Z}) \rightarrow 0 \label{short7}
\;\;.
\ee

Essentially we learn that (as sets)
\be
\mathrm{H}^{2}(M_{(g,p)}, \mathbb{Z})= \mathrm{H}^{1}(\Sigma_{g}, \mathbb{Z})
\oplus \mathrm{H}^{2}(\Sigma_{g}, \mathbb{Z})/p\mathbb{Z} =
\mathrm{H}^{1}(\Sigma_{g}, \mathbb{Z}) 
\oplus \ZZ_p\;\;, 
\ee
with the second summand arising from pull-back of classes (and hence line
bundles) from the base.
Consequently, finite order bundles on $M_{(g,p)}$ are classified by
$\ZZ_p$, pull-backs of bundles from the base $\Sigma_{g}$ are of finite
order $p$, and all such finite order line bundles on $M_{(g,p)}$ arise
as pull backs of bundles from the base.

For $p=0$, i.e.\ the trivial bundle $\Sigma_g \times S^1$, 
the Gysin sequence gives back the K\"unneth formula,
\be
\mathrm{H}^{2}(M, \mathbb{Z}) 
= \mathrm{H}^{2}(\Sigma_{g} , \mathbb{Z}) \oplus
\mathrm{H}^{1}(\Sigma_{g}, \mathbb{Z}) . \mathrm{H}^{1}(S^{1}, \mathbb{Z})
= \mathrm{H}^{2}(\Sigma_{g} , \mathbb{Z}) \oplus
\mathrm{H}^{1}(\Sigma_{g}, \mathbb{Z})\;\;,
\ee
there is no torsion, and line bundles on the base and their
pull-backs are both classified by $\mathrm{H}^{2}(\Sigma_{g} , \mathbb{Z})$.

\subs{Diagonalisation}

Let $P_{G}$ be a principal $G$ bundle over $\Sigma_{g}$ and denote by
$\pi_V: V_{\lg}\ra \Sigma_g$ the associated $\mathbf{ad}$ vector bundle.
For $G$ simply
connected, $P_G$ and  $V_{\lg}$  are necessarily trivial(isable). 
Let $\varphi$ be a section of $V_\lg$. As explained in
\cite{btg/g,btdiag}, one may conjugate $\varphi$ (by vertical automorphisms
$g \in \mathrm{Map}(M,G)$ of $V_{\lg}$) such that it takes values
in the Cartan subalgebra $\lt$ of the Lie algebra $\lg$ of $G$. As  such,
$\varphi$ becomes a section of a $\lt$-bundle $V_{\lt}$ over $\Sigma_g$
which need however not be trivial.  $\lt$-bundles are classified by 
$\mathrm{H}^2(\Sigma_g,\ZZ^{\mathbf{rk}})\sim
\ZZ^{\mathbf{rk}}$, and all $\lt$-bundles are engendered in this way. In
other words, all non-trivial $\lt$- or $T$-bundles arise as obstruction 
bundles to diagonalisation in this case.

In general, diagonalisation and the obstructions to diagonalisation
are not well understood in more than 2 dimensions (see the discussion in
\cite{btdiag}). However, all we need is (see section 2) the diagonalisation
of $U(1)$-invariant $\lg$-valued sections on $M_{(g,p)}$ and this is
straightforward.

First of all, we can pull back $V_{\lg}$ to $M_{(g,p)}$, 
$\pi^*V_{\lg} \approx M_{(g,p)}\times \lg$.  The diagonalising maps 
can be pulled back
to $M_{(g,p)}$ as well and so we have diagonalised $\pi^{*}\varphi$ on
$M_{(g,p)}$. If the diagonalised $\varphi$ was a section of $V_{\lt}$,
$\pi^*\varphi$ is a section of $\pi^*V_{\lt}$. Thus the pull-backs of
$\lt$-bundles on $\Sigma_g$ to $M_{(g,p)}$ are the obstructions to
diagonalising sections of pull-back bundles. 

Secondly, sections of pull-back bundles are in one-to-one correspondence with
$U(1)$-invariant sections of bundles on $M_{(g,p)}$: clearly by definition
$\pi^*\varphi$ is $U(1)$-invariant, and conversely any $U(1)$-invariant
section is such that its value depends on $m\in M_{(g,p)}$ only via $\pi(m)
\in \Sigma_g$, and hence it can be identified with the section of a line
bundle on the base.

We can thus conclude that the problem of diagonalising $U(1)$-invariant 
$\lg$-valued sections of bundles on $M_{(g,p)}$ can be reduced to the
well-understood problem of diagonalisation on the base $\Sigma_g$. The
diagonalised sections are sections of the pull-backs of the $\lt$-bundles
$V_{\lt}$ on $\Sigma_g$. As we have shown, these are precisely the torsion
bundles on $M_{(g,p)}$. Thus the price of conjugating an invariant section 
into the Cartan subalgebra $\lt$ is to 
``liberate'' all these $\lt$-bundless of finite order on $M_{(g,p)}$.
This accounts for the ubiquitous appearance of summations over 
$\ZZ_p^{\mathbf{rk}}$ in the equations of sections 5 to 7.

\section{Regularising the Determinants}\label{regs}

The path integrals that we encounter in the text are formally
roots of determinants,
\be
\int D \Phi \; \exp{ \left(i \int \Phi * Q \, \Phi \right)} =
\frac{1}{ \sqrt{\Det{Q}}}\;\;. 
\label{phidet}
\ee
Because of the oscillatory nature of the path
integral, we set, for an operator $Q$,
\be
\sqrt{\Det{Q}} = \sqrt{\left|\Det{Q}\right|}\, \exp{\frac{+i\pi}{2} \,
  \eta(Q)}
\ee
where
\be
\eta(Q) = \frac{1}{2} \sum_{\lambda} \mathrm{sign}(\lambda)
\ee
and $\lambda$ are the eigenvalues of $Q$ and the root is the positive
root. The $+$ sign in front of $\eta(Q)$ in the phase appears here because
the scalar product (trace) implicit in (\ref{phidet}) is negative definite 
for anti-hermitian fields.

Both the absolute value and
the phase of the determinants require regularisation. We
regularise the absolute value and the phase (assuming that zero is not
an eigenvalue) by setting
\bea
\left|\Det{Q}\right|(s) &=& \exp{\sum_{\lambda} e^{ s \Delta} \ln
  |\lambda|} \label{absdet}
\\
\eta(Q, \, s) &=& \frac{1}{2} \sum_{\lambda}
\frac{\mathrm{sign}(\lambda)}{|\lambda|^{s} }\exp{s \Delta} \label{etas}
\eea
for some appropriate operator $\Delta$.

\subs{Some Geometry}

As everything in
sight respects the geometric $U(1)$ structure on $M$, we decompose all
fields with respect to this action. This means that, as in \cite[section
5]{BW} we expand all the
fields in eigenmodes of the Lie derivative $L_{K}$. So we
set
\be
A_{H} = \sum_{n=- \infty}^{\infty} A_{n},
\ee
where the eigenmodes satisfy
\be
L_{K}\, A_{n}= - 2\pi i n\,  A_{n} \;\;\; \iota_{K}\, A_{n} = 0\;\;,
\ee
and likewiswe for the ghosts $c$ and $\overline{c}$.
By the discussion in Appendix \ref{apgysin}, these eigenmodes can
equivalently be regarded as sections of line bundles $\mathcal{O}(-np)$
over $\Sigma_g$ (which pull back to the trivial line bundle on
$M_{(g,p)}$).

Hence we have that
\be
\Omega^{0}(M, \mathbb{C}) = \bigoplus_{n}\, \Omega^{0}(
\Sigma_{g},\mathcal{O}(-np) ),
\ee
and on tensoring with the trivial bundles $V_{\lk}$ below and 
$\pi^{*}(V_{\lk}) = M \times \lk$ above we have
\be
\Omega^{0}(M, \lk) = \bigoplus_{n}\, \Omega^{0}( \Sigma_{g},
\mathcal{O}(-np)\otimes V_{\lk})\;\;.\label{f2} 
\ee
A similar discussion shows that each mode $n$ of a horizontal 1-form on
$M$ is one to one with a section on $\Sigma_{g}$, consequently one has
\be
\Omega^{1}_{H}(M, \lk) = \bigoplus_{n} \, 
\Omega^{1}( \Sigma_{g}, \mathcal{O}(-np)\otimes V_{\lk}). \label{f1}
\ee

\subs{Computing the Ratio of Determinants}

As explained in \cite{btg/g}, ratios of determinants of the form of
(\ref{ratio}) almost cancel. They would cancel except for mismatches
of harmonic modes. 
The ratio of determinants (\ref{ratio}) is essentially
\be
\prod_{n}\left(2\pi n + \ad \phi \right)_{\lk}^{\dim \Omega^{0}(\Sigma_{g},
  \mathcal{O}(-np) ) -\frac{1}{2}\dim \Omega^{1}(\Sigma_{g},
  \mathcal{O}(-np) ) }
\ee

The ratio may now be deduced by using the Hodge decomposition and an
index theorem on $\Sigma_{g}$. We set
\be
\Omega^{1}(\Sigma_{g},
\mathcal{O}(-np)) = \Omega^{(0,1)}(\Sigma_{g},
\mathcal{O}(-np)\otimes V_{\lk}) \oplus \Omega^{(1,0)}(\Sigma_{g},
\mathcal{O}(-np)\otimes V_{\lk}) ,
\ee
with the complex structure on $\Sigma_{g}$ chosen to be consistent
with the standard orientation, $*_{\Sigma} \, dz = -i \, dz$.

Consider the determinant coming from the ghosts. This is
\be
| \Det{i(\mathcal{L}_{\phi})}| 
\ee
acting on $\Omega^{0}(\Sigma_{g}, \mathcal{O}(-np)\otimes V_{\lk})$
and the absolute value is there, since the ghost determinant should be a
real volume (of the gauge group). Using our regularisation we re-write
the ghost determinant as
\be
\sqrt{\Det{i(\mathcal{L}_{\phi})}}.\sqrt{\Det{i(\mathcal{-L}_{\phi})}}
\ee
the phases cancelling between the square roots since, by (\ref{etas}),
$\eta(Q,s)+ \eta(-Q,s)=0 $.

Upon decomposing the space of 1-forms as
\be
\Omega^{1}(\Sigma_{g}, \mathcal{O}(-np)) = \Omega^{(1,0)}(\Sigma_{g},
\mathcal{O}(-np)) \oplus \Omega^{(0,1)}(\Sigma_{g},
\mathcal{O}(-np))
\ee
the ratio of determinants (\ref{ratio}) that we wish to calculate becomes
\be
\prod_{n=- \infty}^{\infty} \sqrt{\frac{\Det{\left(i\mathcal{L}_{ \phi}
      \right)_{\Omega^{0}(\Sigma_{g}, \mathcal{O}(-np)\otimes V_{\lk})} 
    }}{\Det{\left(i\mathcal{L}_{ \phi}
      \right)_{\Omega^{(1,0)}(\Sigma_{g}, \mathcal{O}(-np)\otimes V_{\lk} )} 
    }}\, . \,\frac{\Det{\left(-i\mathcal{L}_{ \phi}
      \right)_{\Omega^{0}(\Sigma_{g}, \mathcal{O}(-np)\otimes V_{\lk})} 
    }}{\Det{\left(-i\mathcal{L}_{ \phi}
      \right)_{\Omega^{(0,1)}(\Sigma_{g}, \mathcal{O}(-np)\otimes V_{\lk})} 
    }} } \;\;.\label{Ratio}
\ee

Since we are calculating ratios of determinants the traces in both
(\ref{absdet}) and (\ref{etas}) will involve the differences of traces
on the spaces $\Omega^{*}(\Sigma_{g}, \mathcal{O}(-np)\otimes
V_{\lk})$. The eigenvalues are constants, so that the regularised
traces can be evaluated directly,  as $s \rightarrow 0$, on setting
$\Delta$ to be the appropriate Laplacians
\be
[\Tr_{\Omega^{(0,0)}(\Sigma_{g}, \mathcal{O}(-np)\otimes V_{\lk})} -
\Tr_{\Omega^{(0,1)}(\Sigma_{g}, \mathcal{O}(-np)\otimes
  V_{\lk})}] \exp{s \Delta} = \chi(\mathcal{O}(-np)\otimes V_{\lk})
\ee
and
\be
 [\Tr_{\Omega^{(0,0)}(\Sigma_{g}, \mathcal{O}(-np)\otimes
  V_{\lk})} - 
\Tr_{\Omega^{(1,0)}(\Sigma_{g}, \mathcal{O}(-np)\otimes
  V_{\lk})}]\exp{s \Delta} = - \chi(K \otimes
\mathcal{O}(-np)\otimes V_{\lk})\;\;, 
\ee
where $K$ is the canonical bundle of $\Sigma_{g}$.
Note that
\bea
\chi(\mathcal{O}(-np)\otimes V_{\lk}) + \chi(K \otimes
\mathcal{O}(-np)\otimes V_{\lk}) & = & 2\left(
  c_{1}(\mathcal{O}(-np)) + c_{1}(V_{\lk}) \right) \nonumber \\
\chi(\mathcal{O}(-np)\otimes V_{\lk}) - \chi(K \otimes
\mathcal{O}(-np)\otimes V_{\lk}) & = &  \chi(\Sigma_{g}) .
\eea

We decompose the complexified Lie algebra as $\lg_{\mathbb{C}}
= \lt_{\mathbb{C}} \oplus \lk_{\mathbb{C}}$. Denote the roots by
$\alpha$. Furthermore we decompose $\lk_{\mathbb{C}}$ in terms of root
spaces and we write $V_{\lk} = \oplus_{\alpha} V_{\alpha}$.  $\ad \phi
$ acts on $V_{\alpha}$ by multiplication by $\alpha(\phi)$.  With our
conventions $i \alpha(\phi)$ is real.
Covariant derivatives on the root space $V_{\alpha}$ are $d + \ad A =
d + \alpha(A)$. Hence $c_{1}(V_{\alpha}) = -\alpha(F_{A})/2\pi i$. 
Moreover, for $X$, $Y \in \lt$ we have
\be
\Tr \ad(X) \, \ad(Y) = - 2\sum_{\alpha >0} i \alpha(X)\,
i\alpha(Y) = 2 c_{\lg} \Tr X\, Y  \label{CK}
\ee
with $c_{\lg}$ the Coxeter number.

\subs{The Absolute Value}

Set for any field $\psi_{n\alpha} \in \Omega^{*}(\Sigma_{g},
\mathcal{O}(-np)\otimes V_{\alpha})$, 
\be
i \mathcal{L}_{\phi} \, \psi_{n\alpha} = M_{n, \, \alpha}\, 
\psi_{n\alpha}, \;\;\; 
M_{n, \, \alpha} = (2\pi n + i\alpha(\phi))
\ee
so that the determinants that we are interested in are essentially
products of such $M_{n,\; \alpha}$ over the roots,
\be
\Det{\left(i\mathcal{L}_{ \phi}
      \right)_{\Omega^{*}(\Sigma_{g}, \mathcal{O}(-np)\otimes V_{\lk})} 
    } = \prod_{\alpha} \Det_{\Omega^{*}(\Sigma_{g}, \mathcal{O}(-np))}
    M_{n, \; \alpha}
\ee
Thus the absolute value of the ratio of determinants is
\bea
& & \left| \frac{\Det{\left(i\mathcal{L}_{ \phi}
      \right)_{\Omega^{0}(\Sigma_{g}, \mathcal{O}(-np)\otimes V_{\lk})} 
    }}{\Det{\left(i\mathcal{L}_{ \phi}
      \right)_{\Omega^{(0,1)}(\Sigma_{g}, \mathcal{O}(-np)\otimes V_{\lk})} 
    }}
\frac{\Det{\left(-i\mathcal{L}_{ \phi}
      \right)_{\Omega^{0}(\Sigma_{g}, \mathcal{O}(-np)\otimes V_{\lk})} 
    }}{\Det{\left(-i\mathcal{L}_{ \phi}
      \right)_{\Omega^{(1,0)}(\Sigma_{g}, \mathcal{O}(-np)\otimes V_{\lk})} 
    }} 
\right|\nonumber \\
& & \;\;\;  = \exp{ \sum_{n, \; \alpha} \left(
    \chi(\mathcal{O}(-np)\otimes V_{\lk}) - \chi(K \otimes 
\mathcal{O}(-np)\otimes V_{\lk} ) \right) \log M_{n,
    \alpha}} \nonumber \\
& & \;\;\;  = \exp{ \sum_{n, \; \alpha} \chi(\Sigma_{g})  \log M_{n,
    \alpha}} = \prod_{\alpha} \prod_{n} \left( 2\pi n + i \alpha(\phi)
\right)^{\chi(\Sigma_{g})}\;\;.
\eea
Here we recognise the infinite product representation of $\sin^2
i\alpha(\phi)/2$. Comparing with (\ref{ts1}),
\be
T_{S^{1}}(\phi) = \det{}_{\lk}{(1- \Ad{ \, \ex{\phi}})}
=\prod_{\alpha>0}(1-\ex{\alpha(\phi)})(1-\ex{-\alpha(\phi)})
\ee
we deduce that 
\be
\left| \frac{\Det{\left(i\mathcal{L}_{ \phi}
      \right)_{\Omega^{0}(\Sigma_{g}, \mathcal{O}(-np)\otimes V_{\lk})} 
    }}{\Det{\left(i\mathcal{L}_{ \phi}
      \right)_{\Omega^{(0,1)}(\Sigma_{g}, \mathcal{O}(-np)\otimes V_{\lk})} 
    }}
\frac{\Det{\left(-i\mathcal{L}_{ \phi}
      \right)_{\Omega^{0}(\Sigma_{g}, \mathcal{O}(-np)\otimes V_{\lk})} 
    }}{\Det{\left(-i\mathcal{L}_{ \phi}
      \right)_{\Omega^{(1,0)}(\Sigma_{g}, \mathcal{O}(-np)\otimes V_{\lk})} 
    }} 
\right|
= \mathcal{N}\,  T_{S^{1}}(\phi)^{\chi(\Sigma_{g})}
\ee
where $\mathcal{N}$ is one of the constants that we do
not need to keep track of since we are independently normalising the
path integral.

\subs{The Phase}

Now we come to the calculation of the phase of the determinant, in
particular the shift in the level $k$. Formally, the calculation is
similar to an analogous calculation in \cite{BW}. However, due to our
gauge fixing, as opposed to non-Abelian localisation, and because we
also need to find the correction to the $\phi^2$-term in the action,
some of the details of the calculation are quite different.

The phase of the products of the ratios of determinants in (\ref{Ratio})
is the sum of the phases, i.e.
\bea
\eta_{\phi}(s)&= & \eta_{(0,1)}(i\mathcal{L}_{\phi})(s) +
\eta_{(1,0)}(-i\mathcal{L}_{\phi})(s) \nonumber \\
& =& -\frac{1}{2}\sum_{n , \; \alpha } \left( \chi(\mathcal{O}(-np)\otimes
  V_{\alpha}) + 
  \chi(K \otimes  
\mathcal{O}(-np)\otimes V_{\alpha} ) \right)  \frac{\mathrm{sign}(2\pi
n+
i\alpha(\phi))}{ |2\pi n+i\alpha(\phi)|^{s}}\nonumber \\
& =&- \sum_{n , \; \alpha }\left(
  c_{1}(\mathcal{O}(-np)) + c_{1}(V_{\alpha})
\right)\frac{\mathrm{sign}(2\pi n+
i\alpha(\phi))}{ |2\pi n+i\alpha(\phi)|^{s}}
\ee

Without loss of generality we choose $\phi$ such that 
$0 <i \alpha(\phi) < 2\pi$ for the positive roots, so that
\bea
\eta_{\phi}(s) & =& - 2 \sum_{\alpha >0} c_{1}(V_{\alpha}) |i
\alpha(\phi)|^{-s} - 2\sum_{n\geq 1}\sum_{\alpha >0} 
    (c_{1}(\mathcal{O}(-np)+ c_{1}( V_{\alpha})) (2\pi n+i \alpha(\phi))^{-s}
    \nonumber \\
& & - 2\sum_{n\geq 1}\sum_{\alpha >0}
    (c_{1}(\mathcal{O}(-np) -c_{1}( V_{\alpha})) (2\pi n- i \alpha(\phi))^{-s}
\eea
The Riemann $\zeta$-function, $\zeta(s) = \sum_{n\geq 1} n^{-s}$,
is regular
on the real axis away from $s=1$, satisfies $\zeta(-1) = -1/12$,
$\zeta(0) = -1/2$  and behaves as 
\be
\zeta(s+1)= \frac{1}{s} + \gamma_{0} +  s\gamma_{1} + \dots
\ee
as $s$ approaches 0. This means that if we expand $(2\pi n\pm i
\alpha(\phi))^{-s}$ in a Taylor series in $\alpha$ we need only keep
terms up to quadratic order in $\alpha(\phi)$. We need to keep
quadratic terms since $c_{1}(\mathcal{O}(-np)) = -np$. Consequently we
have
\bea
\eta_{\phi}(s) & =& - 2 \sum_{\alpha >0} c_{1}(V_{\alpha}) - 
\frac{p}{6} (\dim G
- \dim T) \nonumber \\ 
& & +\frac{2}{\pi} \sum_{\alpha >0} c_{1}(V_{\alpha})\, i\alpha(\phi)
+ \frac{p}{2\pi^{2}} 
\sum_{\alpha >0} i\alpha(\phi) \, i\alpha(\phi) + \mathcal{O}(s)
\eea
As $s \rightarrow 0$ all other terms vanish. 

The other ratio of determinants (\ref{ratio2}) has a phase which
depends neither on $\phi$ nor on the connection and the constant mode
is not included. So following through
the same steps, there is now 
no sum over $\alpha$, and we see that the corresponding 
$\eta$-invariant is
\bea
\eta_0(s) &=& -\dim T \sum_{n \neq 0} c_{1}(\mathcal{O}(-np)) 
\frac{\mathrm{sign}\,
  2\pi n}{|2\pi 
  n|^{s}}  = 2p \dim T \sum_{n>0} n^{1-s}
\non
&\stackrel{s\ra 0}{\ra}& -\frac{1}{6} p\, \dim T\;\;.
\eea

Putting the pieces together, making use of (\ref{CK}) and rewriting the
terms involving the fields $\phi$ and $F_A$ as integrals over $\Sigma_g$,
we find that
the phase of the ratio of the (square roots of) determinants
(\ref{ratio}, \ref{ratio2}) is
\be
-\frac{i\pi}{2}(\eta_{\phi}(0) + \eta_0(0)) = 
\frac{i c_{\lg}}{4\pi}
\int_{\Sigma_g}\left(p \Tr\phi^2 + 2 \Tr \phi F_A\right)
 + 4\pi i p \Phi_0 - \int_{\Sigma_g}\rho(F_H) 
\ee
The first term corresponds to a shift of the level of the action,
\be
\frac{k}{4\pi}\int_{\Sigma_g} \left(p \Tr\phi^2 \omega + 2 \Tr \phi F_A\right)
\ra
\frac{k+c_{\lg}}{4\pi}\int_{\Sigma_g} 
\left(p \Tr\phi^2 \omega + 2 \Tr \phi F_A\right)
\ee
The second term is an overall $p$-dependent phase, with 
\be
\Phi_0 = \frac{1}{48}\dim G\;\;,
\ee
whose significance (framing) we discuss in section 8. Finally, the third
term, which arises from
\be
-i\pi \sum_{\alpha>0} c_1(V_\alpha) =
-\int_{\Sigma_g}\frac{1}{2}\sum_{\alpha>0} \alpha(F_H) \equiv
-\int_{\Sigma_g} \rho(F_H) \;\;,
\ee
makes no contribution for simply-connected groups for which $\rho(F_H)$ 
is an integral multiple of $2\pi i$. 
However, for non-simply connected groups this term is potentially
non-trivial and cannot be neglected (see e.g.\ \cite{bhtkz}).

\section{The Expectation Value of the Hopf Link}\label{othersurg}

We arrived at the expectation value for the Hopf link results by using
(\ref{final2a}) one can reasonably 
ask what we would find if we had used (\ref{final1}) instead. We
consider here the case where $M = S^{3}$ and $G= SU(2)$ so that 
characters of the representations are, (with $\phi$ now the
coefficient in a basis of simple roots)
\be
\chi_{j}(\phi) = \frac{\sin (j+1)\phi}{\sin \phi}.
\ee
A simple calculation, by writing each sine as a difference of phases,
allows us to determine $K$,
\bea
K_{ij}&= & <\chi_{i}(\phi)\chi_{j}(\phi)> \nonumber \\
&= &  \frac{2 }{\pi}\,\ex{4\pi i
  p\Phi_{0}}\, \int_{-\infty}^{\infty} d\phi \,
 \sin{(
(i+1)\phi)}\; \sin{ ((j+1)\phi)} \exp{(-i \frac{k+2}{2\pi}
\phi^{2}) }\nonumber \\
&= &  \ex{2\pi i  (\Phi(i) +\Phi(j) )}\,
S_{ij}\nonumber \\
&=& T(i)\, S_{ij}\, T(j) \label{ST}
\eea
where
\be
S_{ij} =
  \sqrt{\frac{2}{k+2}}\, 
  \sin{\frac{(i+1)(j+1)\pi}{k+2} } \label{ST1}
\ee
and the phase is
\be
\Phi(j) = \frac{c_{2}(j)}{(k+ 2)} - \frac{3}{16} + p\Phi_{0}
+ \frac{1}{4(k+2)}
 =  \frac{c_{2}(j)}{(k+ 2)}- \frac{k}{8(k+2)}\label{frame}
\ee
$c_{2}(j) = j(j+2)/4 $ and $T(j)$ agrees with the formula that was
given in (\ref{T}). We have thus obtained (\ref{K1}) without using
(\ref{maprels}).

\rnc{\Large}{\normalsize}

\end{document}